\documentclass{article}
\usepackage{ffpreprint}
\usepackage[utf8]{inputenc}
\usepackage[T1]{fontenc}
\usepackage{hyperref}
\usepackage{url}
\usepackage{booktabs}
\usepackage{amsfonts}
\usepackage{nicefrac}
\usepackage{microtype}
\usepackage{graphicx}
\usepackage{natbib}
\usepackage{doi}
\usepackage{amsmath}
\usepackage{amssymb}
\usepackage{subcaption}
\usepackage{algorithm}
\usepackage{algpseudocode}

\title{Moment Optimization in the Navascu\'es-Pironio-Ac\'\i n hierarchy}
\author{
  Francesco Flora$^{1,2}$ \quad
  Losel Matos$^{1,4}$ \quad
  Tim Heightman$^{1}$ \quad
  Tamás Kriváchy$^{1,5}$ \quad
  Adan Garriga$^{2}$ \quad
  Antonio Acín$^{1,3}$ \\[6pt]
  {\small $^{1}$ ICFO - Institut de Ciencies Fotoniques, The Barcelona Institute of Science and Technology, 08860 Castelldefels (Barcelona), Spain}\\
  {\small $^{2}$ Eurecat, Centre Tecnològic de Catalunya, Barcelona, Spain} \\
  {\small $^{3}$ ICREA - Institució Catalana de Recerca i Estudis Avan\c cats, Pg. Llu\'\i s Companys 23, 08010 Barcelona, Spain} \\
  {\small $^{4}$ École Polytechnique, Institut Polytechnique de Paris}\\
  {\small $^{5}$ Institute for Theoretical Physics, ETH Zurich, 8093 Zurich, Switzerland}
}
\hypersetup{
  pdftitle={Moment Optimization in the Navascu\'es-Pironio-Ac\'\i n hierarchy},
  pdfsubject={quant-ph},
  pdfauthor={Francesco Flora},
  pdfkeywords={NPA hierarchy, moment selection, Bell inequalities,
               semidefinite programming, Bayesian optimization,
               parallel tempering, restricted Boltzmann machine}
}
\begin{document}
\maketitle
\begin{abstract}
The Navascués--Pironio--Acín (NPA) hierarchy provides a convergent sequence of semidefinite programming (SDP) relaxations for bounding the solution to noncommutative polynomial optimisation problems, ubiquitous in quantum physics.
Its practical applicability is however limited by the computational overhead when increasing the hierarchy level, due to the combinatorial growth in the number of operator moments that must be included at each level. Nevertheless, it is known that not all the moments have the same impact on the quality of the bounds obtained through NPA and it is a relevant problem to understand how to select moments to get tighter bounds for a fixed computational effort.
In this work, we first reframe the problem of choosing moments in NPA relaxations as one
of combinatorial subset selection. Given a computational budget, we show how to select
moments from a candidate pool to achieve the tightest possible bound.
We show that this selection problem is governed by strong higher-order synergistic interactions among moments, which we quantify through a marginal synergy diagnostic adapted from the study of
complex systems. We then develop and compare three
complementary optimization methods for moment selection: Parallel Tempering (PT), deep policy-based reinforcement learning with a Restricted
Boltzmann Machine (RBM) architecture, and Bayesian Optimization (BO).  Using the $I_{3322}$ Bell inequality as a benchmark, we demonstrate that all three methods substantially outperform greedy optimisation approaches at computational costs around two orders of magnitude below brute force, with the RBM achieving the closest approach to optimal bounds throughout the hard transition regime.
We illustrate the power of our framework in two paradigmatic problems.  First, we study all the 174 bipartite Bell inequalities involving four measurements of two outcomes and demonstrate how improved bounds can be obtained through moment selection optimisation.
Second, we move to the certification of ground-state properties
of the one-dimensional Heisenberg spin chain, demonstrating that physically
motivated monomial bases are internally compressible and are not globally optimal in general. Our budget-aware
search over a broader monomial pool improves the certified bound on long-range
correlations by nearly two orders of magnitude.  Together, these results
establish a general, scalable framework for moment selection in noncommutative
polynomial optimization, with applications in many scenarios in quantum physics and quantum information theory.
We provide an open-source implementation of our method, applicable to
general moment optimization problems, at
\url{https://github.com/Ciscone7/OptimalMomentsNPA}.
\end{abstract}
\section{Introduction}
\label{sec:intro}

A recurring task in quantum physics consists of the optimisation of the expectation value of a noncommutative
polynomial over quantum states and operators~\cite{pironio2010convergent}.  Bounding the maximum quantum violation of a Bell
inequality, for instance, or estimating the ground-state energy of a many-body
Hamiltonian, can both be cast in this form.  The
Navascués--Pironio--Acín (NPA) hierarchy~\cite{NPA2007, NPA2008} addresses
such problems by constructing a sequence of semidefinite programming (SDP)
relaxations, each parameterized by a finite set of operator \emph{moments}, that provide a convergent sequence of bounds to the searched optimal solution.
The feasible sets of these relaxations form progressively tighter outer
approximations to the set of feasible points in the optimisation problem, and convergence to the exact quantum value
is guaranteed in the limit.  At any finite level $n$, denoted in this work by $\textsf{NPA}n$, the hierarchy yields a rigorous bound.

The framework has become a standard tool across quantum information science, especially in the device-independent scenario.
It represents the primary numerical method for upper-bounding Bell inequality
quantum violations~\cite{BrunnerEtAl2014, CollinsGisin2004, PalVertesi2010,
tsirelson1987quantum} and for certifying correlations in device-independent
protocols.  It also has been
applied to certify ground-state~\cite{wang2024certifying} and
steady-state~\cite{mortimer2025certifying,robichon2024certifying} properties of many-body quantum systems.
The mathematical foundations connecting positivity constraints on moment matrices
to the operator-algebraic structure of noncommutative polynomials are now well
established~\cite{pironio2010convergent}, and efficient numerical
implementations exist~\cite{wittek2015algorithm}.
In practice, however, the hierarchy is constrained by a computational
bottleneck that limits its reach.  The number of required moments, and therefore the size of the matrices involved in the SDP, scales
combinatorially with the hierarchy level, so that
higher-level relaxations quickly become computationally prohibitive.  The usual
remedy is to work at a fixed, low level and accept the resulting gap to the
true optimum.  This raises a natural question that has thus far received little
attention in the quantum information community: rather than including \emph{all}
moments up to a given level, can one identify which \emph{subset} of moments
yields the tightest relaxation within a fixed computational budget?

In this work, we formalize this question as follows.  We define an initial set of moments
$\mathcal{I}$, for instance the minimal set necessary to define the cost function in the polynomial optimisation problem, and a larger set $\mathcal{F}$ of candidate moments that can be selected. We also define the computational cost by the maximal number $k$ of moments to be chosen from the set $\mathcal{A} = \mathcal{F}\setminus\mathcal{I}$ of size $N$. Each choice of $k$ moments from $\mathcal{A}$ produces a different bound and the goal is to find the choice of these $k$ moments that produce the tightest bound.
This moment optimisation cannot rely on gradients or closed-form structure; it is a black-box
combinatorial optimization problem, and the difficulty of navigating it is what
drives the need for the methods developed in this work.

To grasp an intuition about the structure of the problem, we work in detail on the
$I_{3322}$ Bell inequality~\cite{CollinsGisin2004, BrunnerEtAl2014}, defined in a scenario where two parties, Alice and Bob, each perform three measurements
of 
two outcomes. It represents a standard benchmark with a well-understood
NPA behavior, here used to derive upper bounds to the maximal quantum violation of the inequality.  In this first studied scenario, the initial set is taken equal to the first level of the hierarchy, $\mathcal{I}=\textsf{NPA1}$, and the candidate pool as the second level, $\textsf{NPA2}$, so that $\mathcal{A} = \textsf{NPA2}\setminus\textsf{NPA1}$ contains $N = 21$
candidate moments. The size of the full configuration space of possible moment selections is $2^{21}$, approximately two million subsets within reach of exhaustive enumeration.  This
gives ground-truth optimal solutions at every budget $k$, making $I_{3322}$ an
ideal setting in which to expose and characterize the structure of the
moment-selection landscape before deploying heuristic optimizers.
A brute-force exploration of all the $2^{21}$ subsets reveals that not all moments contribute equally to the tightness of the corresponding upper bound on the maximal quantum violation of $I_{3322}$.
The landscape is in fact highly heterogeneous: the
typical value across random subsets is far from the best achievable value.  The
root cause is strong higher-order synergy among the moments --- the best pair
does not in general contain the best single moment, and the optimal subset of
size~$k$ cannot be built by extending the optimal subset of size $k{-}1$.
Greedy strategies, which evaluate moments individually, miss this collective
structure entirely and fail to improve until a large fraction of the adding set
is included. To quantify these collective effects, we adapt the synergy-first backbone
decomposition of~\cite{varley2024synergy} to introduce the \emph{marginal
synergy} $\Delta(S_k)$, which measures how much the relaxation bound degrades
on average when any single moment is removed from the best known $k$-subset.

The synergistic landscape, reminiscent of spin glasses~\cite{mezard1987spin,
barahona1982computational} with rugged energy landscapes and frustrated local
minima, rules out any strategy based on individual moment assessments.
What is needed is a global search capable of navigating high energy barriers and
identifying collectively important subsets.  We develop three complementary
methods for this: \emph{Parallel Tempering} (PT)~\cite{hukushima1996exchange,
earl2005parallel}, which runs multiple replicas at different temperatures and
uses exchange moves to cross energy barriers; a policy-based reinforcement
learning approach using a \emph{Restricted Boltzmann Machine}
(RBM)~\cite{hinton2002training, goodfellow2016deep} policy; and \emph{Bayesian Optimization} (BO)~\cite{shahriari2016taking}
with a random forest surrogate.  Detailed descriptions of all three methods
are collected in a dedicated section later in the article.
On the $I_{3322}$ benchmark, all methods substantially outperform the
greedy baseline at costs two orders of magnitude below brute
force, with RBM achieving  the closest tracking of the optimal path throughout
the hard transition regime. 

We then apply the developed optimisation framework to two settings where brute-force moment optimisation is infeasible.  For the 174 Bell
inequalities in the $(4,4,2,2)$ scenario, where Alice and Bob perform four measurements of two outcomes, we find that the transition from NPA1 bound to the NPA2 bound begins well below the NPA2 budget, and we find substantial heterogeneity in
the moment budget required to reach a given accuracy, confirming that the coarse
``\textsf{NPA2} bound'' label conflates qualitatively different convergence
regimes. We then explore the impact of moment selection in a many-body context, namely to study ground-state properties of the one-dimensional Heisenberg spin chain. We show that the
physically motivated local basis of Ref.~\cite{wang2024certifying} is
almost optimal yet redundant for the ground-state energy, but not for the long-range correlator:
enlarging the candidate pool and warm-starting from the local-basis solution
improves the certified bound on long-range correlations by nearly two orders of
magnitude.

Before presenting our results, it is worth commenting on related work by Requena et al.~\cite{requena2023certificates}, where deep
reinforcement learning is used to select SDP constraints for bounding many-body
ground-state energies.  Our approach shares that high-level motivation but
differs in several respects. First, we work directly on the NPA moment-matrix
structure rather than generic SDP constraints. Second, we carry out an
information-theoretic analysis of the landscape that informs the design of our
methods.  Third, we compare a broader set of optimization strategies across
multiple computational budgets.  Furthermore, we believe that the computational
overhead of full deep RL is unnecessary for this problem, and that our
lighter-weight methods are better suited to the specific structure of the
moment-selection landscape.  Indeed, our deep RL implementation is policy-based
and does not come with the usual requirements of replay buffers or value
networks that can significantly increase the compute necessary for effective
inference.

The article is organized as follows. Section~\ref{sec:methods} introduces the
NPA hierarchy and formulates the moment-selection problem.
Section~\ref{sec:problem} presents the $I_{3322}$ benchmark, characterizes
the structure of the moment landscape, and introduces the synergy diagnostic.
Section~\ref{sec:opt_methods} describes the three optimization methods
(Parallel Tempering, Restricted Boltzmann Machine, and Bayesian Optimization);
readers primarily interested in our main results may skip this section and proceed
directly to Section~\ref{sec:results}. Section~\ref{sec:results} collects all
numerical results: the comparison of methods on $I_{3322}$, the application to
all 174 Bell inequalities in the $(4,4,2,2)$ scenario, and the extension to
many-body ground-state certification. Section~\ref{sec:conclusion} summarizes
the findings and discusses directions for future work.
\section{Preliminaries}
\label{sec:methods}
\subsection{Moment representation for noncommuting polynomials}
\label{subsec:operator_to_moment}
Many problems in quantum information theory can be written as
the optimisation of the expectation value of a polynomial over noncommuting operators.  Concretely,
we consider

\begin{equation}
    \label{eq:exact_problem}
    \begin{aligned}
        \mathsf{P}_{\rm exact}
        =
        \inf_{H,\, O_1,\ldots,O_m,\, \rho}
        \quad & \operatorname{Tr}\!\left[\rho\, p(O_1,\ldots,O_m)\right] \\
        \text{subject to}\quad
        & \rho\succeq 0 \\
        &\operatorname{Tr}[\rho]=1 \\
        & \operatorname{Tr}[\rho\, q_i(O_1,\ldots,O_m)] \geq 0
          \quad i = 1,\ldots,I \\
        & r_j(O_1,\ldots,O_m) \succeq 0
          \quad j = 1,\ldots,J,
    \end{aligned}
\end{equation}
where $O_1,\ldots,O_m\in\mathcal{B}(H)$ are bounded operators acting on a Hilbert space $H$ of arbitrary dimension, $\rho$ is a quantum
state on $H$, $p$ is a fixed Hermitian polynomial encoding the objective, and
$\{q_i\}$ and $\{r_j\}$  are polynomials encoding problem-specific algebraic relations that the
operators must satisfy~\cite{pironio2010convergent}.
In its general formulation, the state and operators act on an arbitrary Hilbert space, but we can also consider variants of the problem where the operators are fixed and the optimisation is only over the state. 
In these applications (e.g.\ ground-state
certification), the infimum reduces to $\inf_\rho\operatorname{Tr}[\rho\,p]$
with the $O_i$ removed from the optimisation.

We now introduce the free $\ast$-algebra of noncommuting words in the
generators $O_1,\ldots,O_m$, which provides a systematic language for
indexing these expectation values as \emph{moments} and for translating
each condition defined by $\{q_i\}$ and $\{r_j\}$ into an explicit linear constraint on them.

Let $\langle \mathcal{O} \rangle$ denote the free $\ast$-algebra generated by the
formal symbols
\[
\mathcal{O} = \{O_1,\ldots,O_m\}.
\]
Its elements are finite complex linear combinations of words of the form
\[
w = X_1 X_2 \cdots X_k,
\qquad
X_j \in \{O_1,\ldots,O_m, O_1^\dagger,\ldots,O_m^\dagger\},
\]
together with the empty word $1$, which represents the identity.
Note that we adopt the physics notation $\dagger$ to denote the $\ast$-operation. Its action on words is such that
\[
(O_{i_1}O_{i_2}\cdots O_{i_k})^\dagger
= O_{i_k}^\dagger \cdots O_{i_1}^\dagger,
\]
and extends anti-linearly.
No algebraic relations (such as commutation, projector, or normalization identities)
are imposed at this stage; $\langle \mathcal{O} \rangle$ consists of all
noncommuting words in the generators and their adjoints.

Any noncommuting polynomial $p$ may be expanded as
\begin{equation}
    \label{eq:poly_expansion}
    p(O_1,\ldots,O_m)=\sum_{w} c_w\, w,
\end{equation}
so the objective in~\eqref{eq:exact_problem} becomes
\begin{equation}
    \label{eq:true_moments_objective}
    \operatorname{Tr}[\rho\, p]
    = \sum_{w} c_w\, m_w,
\end{equation}
where we define the \emph{true moments}
\begin{equation}
    \label{eq:true_moment_def}
    m_w := \operatorname{Tr}[\rho\, w],
    \qquad w\in\langle\mathcal{O}\rangle.
\end{equation}
These are the natural variables that appear in the exact formulation of the
optimization problem.

To encode the quantum constraints on $\rho$ in terms of moments, Navascués,
Pironio and Acín~\cite{NPA2007,NPA2008} introduced the (infinite) \emph{moment
matrix}
\begin{equation}
    \label{eq:moment_matrix_def}
    \Gamma_{u,v}
    :=
    \operatorname{Tr}[\rho\,u^\dagger v]
    = m_{u^\dagger v},
    \qquad u,v\in\langle\mathcal{O}\rangle.
\end{equation}
Their key result is that the quantum conditions
\[
\rho\succeq 0, \qquad \operatorname{Tr}[\rho]=1,
\]
are equivalent to the following constraints on $\Gamma$:
\begin{equation}
    \label{eq:moment_constraints}
    \Gamma \succeq 0,
    \qquad
    \Gamma_{1,1}=1.
\end{equation}
The trace-one condition arises because the identity operator corresponds to the
empty word, yielding
\[
\Gamma_{1,1} = \operatorname{Tr}[\rho\,\mathbb{I}] = 1.
\]
For many relevant problems in quantum physics, and in particular for those considered in this work, the operator constraints have the form of equalities, such as projector conditions, completeness, or commutation relations. In this case the operators satisfy

\[
r := \sum_i \alpha_i w_i = 0,
\]
so that for all words $u,v$ one has
\[
\operatorname{Tr}[\rho\, u^\dagger r\, v] = 0
\quad\Longrightarrow\quad
\sum_i \alpha_i\, \Gamma_{u,\, w_i v} = 0,
\]
so every operator identity becomes a linear equality among moment-matrix entries.
For instance, if two operators commute, $[O_i, O_j] = O_i O_j - O_j O_i = 0$,
this imposes the linear constraint
$\Gamma_{u,\, O_i O_j v} = \Gamma_{u,\, O_j O_i v}$ for all words $u, v$.

Similarly, scalar inequality constraints of the form $\operatorname{Tr}[\rho\, q_i] \geq 0$,
with $q_i = \sum_w c_w^{(i)}\, w$, translate by linearity of the trace into a linear
inequality on the moments:
\[
\operatorname{Tr}[\rho\, q_i] = \sum_w c_w^{(i)}\, m_w \geq 0.
\]
For the interested reader, a detailed analysis of the relaxation of 
Eq.~\eqref{eq:exact_problem} can be found in~\cite{pironio2010convergent}.

Using the moments and the moment matrix, the optimization
problem~\eqref{eq:exact_problem} can be rewritten exactly as
\begin{equation}
    \label{eq:exact_moment_formulation}
    \begin{aligned}
        \mathsf{P}_{\rm exact}
        =
        \inf_{\{m_w\}}
        \quad & \sum_{w} c_w\, m_w \\
        \text{subject to}\quad
        & \Gamma \succeq 0 \\
        & \Gamma_{1,1}=1 \\
        & \Gamma \text{ satisfies the linear constraints induced by } \{q_i\} \text{and} \{r_j\}.
    \end{aligned}
\end{equation}
This formulation is fully equivalent to~\eqref{eq:exact_problem}, provided that
$\Gamma$ contains the moments associated with \emph{all} words in
$\langle\mathcal{O}\rangle$.
\subsection{Semidefinite relaxations: the NPA hierarchy}
\label{subsec:SDP_relaxation}
The exact moment formulation~\eqref{eq:exact_moment_formulation} involves the (possibly)
infinite moment matrix $\Gamma$ indexed by all words in $\langle \mathcal{O}
\rangle$ and is therefore computationally intractable.
In two subsequent works~\cite{NPA2008, pironio2010convergent}, the NPA introduced a systematic
sequence of semidefinite relaxations, now known as the \emph{NPA hierarchy}.
Each level restricts the moment matrix $\Gamma$ to the moments associated with
words of length at most $n$ (the \emph{truncation order}), yielding a
finite-dimensional SDP.
These truncated constraints form an increasing sequence of necessary conditions,
each strictly stronger than the previous one, and the associated bounds converge
to the exact quantum value as $n \to \infty$.
In the literature, the standard levels are denoted \textsf{NPA1}, \textsf{NPA2},
\textsf{NPA3}, \emph{etc.}, where, for example, \textsf{NPA2} includes all words
of length $\le 2$ and therefore strictly contains \textsf{NPA1}.
At any finite level, the feasible set of truncated moment matrices constitutes an
\emph{outer approximation} of the set of quantum-realizable moment sequences. That is, every physical moment matrix satisfies the truncated constraints, but the
converse need not hold.
Hence each relaxation optimizes over a superset of the true quantum set,
providing a certified \emph{lower bound} for the minimization
problem~\eqref{eq:exact_problem}.
As the truncation order increases, these outer approximations become tighter and
the bounds improve monotonically. However, this comes at the price of a rapidly
increasing computational cost, since the size of the moment matrix grows
combinatorially with~$n$.

For an integer $n\ge 1$, let
\[
\mathcal{W}_n := \{\, w\in\langle\mathcal{O}\rangle \;:\; |w|\le n \,\}
\]
denote the set of all words in the generators and their adjoints of length at
most~$n$.
The corresponding \emph{truncated moment matrix} is defined by
\begin{equation}
    \label{eq:truncated_moment_matrix}
    \Gamma^{(n)}_{u,v} := \operatorname{Tr}[\rho\,u^\dagger v],
    \qquad u,v\in\mathcal{W}_n.
\end{equation}
This matrix contains all moments $m_{u^\dagger v}$ with $u^\dagger v$ of length
at most~$2n$, and is finite-dimensional for every fixed~$n$.

The truncated moment matrix must satisfy the following necessary conditions:
\begin{itemize}
    \item[\(\bullet\)] \emph{Positivity:}
    \[
    \Gamma^{(n)} \succeq 0.
    \]
    \item[\(\bullet\)] \emph{Normalization:}
    since the identity belongs to $\mathcal{W}_n$, we impose
    \[
    \Gamma^{(n)}_{1,1} = 1.
    \]
    \item[\(\bullet\)] \emph{Truncated algebraic constraints:}
    for every operator identity
    \( r = \sum_i \alpha_i w_i = 0\),
    we impose
    \[
    \sum_i \alpha_i\, \Gamma^{(n)}_{u,\, w_i v} = 0
    \qquad
    \text{whenever } u,w_i v \in\mathcal{W}_n.
    \]
    \item[\(\bullet\)] \emph{Truncated scalar inequalities:}
    for every constraint $\operatorname{Tr}[\rho\, q_i] \geq 0$ with
    $q_i = \sum_w c_w^{(i)}\, w$,
    we impose
    \[
    \sum_w c_w^{(i)}\, \Gamma^{(n)}_{1,w} \geq 0
    \qquad
    \text{whenever all words in the support of } q_i \text{ belong to } \mathcal{W}_n.
    \]
\end{itemize}
These constraints are necessary for $\Gamma^{(n)}$ to arise from a quantum state,
but in general they are not sufficient; thus the feasible region constitutes an
outer approximation of the quantum set.

Restricting the polynomial $p$ to the moments represented in $\Gamma^{(n)}$,
the level-$n$ relaxation of problem~\eqref{eq:exact_problem} takes the form
\begin{equation}
\label{eq:relaxed_problem}
\begin{aligned}
    \mathsf{P}^{(n)}
    =
    \min_{\Gamma^{(n)}}
    \quad & \sum_{w} c_w\, m_w
    \\
    \text{subject to}\quad
    & \Gamma^{(n)} \succeq 0, \\
    & \Gamma^{(n)}_{1,1} = 1, \\
    & \Gamma^{(n)} \text{ satisfies all truncated constraints induced by } \{q_i\} \text{ and } \{r_j\}.
\end{aligned}
\end{equation}
This is a semidefinite program of size $|\mathcal{W}_n|\times|\mathcal{W}_n|$,
typically referred to as \textsf{NPA$n$}.
Since the feasible region of~\eqref{eq:relaxed_problem} strictly contains the set
of quantum-realizable moment matrices, we have
\[
\mathsf{P}^{(1)} \le \mathsf{P}^{(2)} \le \cdots \le \mathsf{P}_{\rm exact},
\]
so each relaxation provides a certified \emph{lower bound} on the true optimum.
In Bell scenarios, for historical reasons the optimisation is posed as a maximisation problem, as one is interested in the maximal quantum violation of a given Bell inequality. Optimizing under these constraints yields numerical upper bounds on maximal quantum violations~\cite{tsirelson1987quantum,NPA2007}.  Similarly, in quantum many-body physics, quantities such as ground-state  and steady-state observables can be cast in terms of noncommutative polynomial optimisation.  This approach allows for the derivation of rigorous bounds to
observables.

The dimension of $\Gamma^{(n)}$ increases rapidly with~$n$, since
$|\mathcal{W}_n|$ grows combinatorially with the number of generators and with
the truncation order.
Thus higher levels offer tighter approximations at the expense of significantly
increased computational cost, and practical applications typically rely on
relatively small values of~$n$.
It is crucial to recognize that the primary computational bottleneck arises from
the rapid growth of the number of operators—hereafter called \emph{moments}—
included at higher hierarchy levels.
The number of moments scales combinatorially (approximately exponentially) with
the hierarchy order, quickly rendering higher levels computationally prohibitive.
Thus, careful management of computational resources and problem-specific
optimizations are essential for efficient practical applications
\cite{NPA2008,wang2024certifying}.
Moreover, within a given level of the hierarchy, not all moments contribute
equally to the quality of the bound.  It is therefore important to develop
efficient tools that, within a finite computational budget, can identify the
most meaningful moments in order to obtain the tightest possible bound given the
available resources.

Although the NPA hierarchy arises from quantum expectation values, the framework
developed above applies to arbitrary noncommutative polynomial optimization
problems.
Evaluating a polynomial $p$ on a state is equivalent to applying a positive,
normalized linear functional $\phi$ on the $\ast$-algebra generated by the
variables.
By the Gelfand--Naimark--Segal (GNS) construction, every such functional admits a
Hilbert-space representation of the form
$\phi(w)=\langle \Omega_\phi,\, \pi_\phi(w)\, \Omega_\phi\rangle$
for a suitable representation $\pi_\phi$ and vector $\Omega_\phi$~\cite{bratteli1987operator}.
Thus any noncommutative polynomial optimization task can be written in
the moment-matrix form used here, with the quantum mechanical formulation
appearing simply as one concrete realization of this general setting.
The commutative analogue of this approach is the Lasserre hierarchy~\cite{lasserre2001global}.
The moment selection methods developed in this work extend naturally also to this commutative case.

\section{Statement of the problem}
\label{sec:problem}

We model the task as a constrained optimization over subsets of moments used to
construct the moment matrix.
Let $\mathcal{I}$ denote an \emph{initial set} of moments and $\mathcal{F}$ a
\emph{final set}, such that
\begin{equation}
    \mathcal{I} \subseteq \mathcal{F}.
\end{equation}
The optimization is performed over the \emph{adding set}, defined as the set
difference
\begin{equation}
    \mathcal{A} := \mathcal{F} \setminus \mathcal{I},
\end{equation}
with $N := |\mathcal{A}|$ denoting its size.
Intuitively, $\mathcal{A}$ represents the collection of candidate moments that
may be added, subject to computational limitations in solving the associated
semidefinite program.
While in many instances the sets $\mathcal{I}$ and $\mathcal{F}$ correspond to
successive levels of the NPA hierarchy, this correspondence is not required in
general.
Formally, each choice $\mathcal{C} \subseteq \mathcal{A}$ is a finite list of
moments,
\begin{equation}
    \mathcal{C} = \{ m_1, m_2, \ldots, m_{|\mathcal{C}|} \},
\end{equation}
where each $m_i$ represents a product of operators in the SDP problem under study.
From any such set $\mathcal{C}$, one can construct the associated moment matrix
$M(\mathcal{C})$, which defines the corresponding relaxation.

With these definitions in mind, we identify a \textit{relaxation} with a binary
vector $x$,
\begin{equation}
    x \in S_N := \{0,1\}^N,
\end{equation}
where the vector $x$ specifies which moments, among those available in
$\mathcal{A}$, are included in a particular relaxation on top of the minimal set $\mathcal{I}$.
This representation allows us to be agnostic to the underlying physics and
naturally encode any symmetries of the problem.
In this notation,
\begin{equation}
    \mathbf{0} \equiv \mathcal{I},
    \qquad
    \mathbf{1} \equiv \mathcal{F},
\end{equation}
where $\mathbf{0},\mathbf{1}\in S_N$ denote the all-zero and all-one vectors,
respectively.
The space $S_N$ can be partitioned into layers of fixed Hamming weight,
\[
    S_N =  \bigsqcup_{k=0}^{N} S_N^k,
\]
where $S_N^k$ is the set of binary vectors of length $N$ with exactly $k$ ones.
We denote elements of $S_N^k$ by $x^{k}$, with
\begin{equation}
    \| x^k \|_1 = k.
\end{equation}
Given a relaxation choice $x$, we solve the associated SDP and obtain a real
number.
Writing $\gamma$ for the parameters that specify the physical problem (for
instance, the coefficients of a Bell inequality), we have
\begin{equation}
    x \in S_N \;\mapsto\; f_{\gamma}(x) \in  \mathbb{R},
\end{equation}
where we interpret $f_{\gamma}$ as the \emph{cost function} of the problem.
With these definitions, our problem of optimal moment selection becomes a combinatorial optimization problem. Indeed, we seek a vector $\hat{x}^k$ such that
\begin{equation}
\label{eq:opt_problem}
    \hat{x}^k
    =
    \arg\min_{x^k \in S_N^k} \; f_{\gamma}(x^k),
\end{equation}
where the optimal solution depends on the problem of interest (e.g., a Bell inequality or a many-body Hamiltonian) encoded in $\gamma$, the adding set $\mathcal{A}$, and the number of moments, $k$, we are allowed to select, which depends on the available computational resources.
Modifying the initial and final sets while preserving the size of their set
difference $N$ alters the optimization landscape.
The goal is to identify the $k$ moments in $\mathcal{A}$ that produce the tightest bound.

The optimization problem considered here—minimizing a function $f_\gamma(x)$ over
binary vectors $x \in \{0,1\}^N$ with fixed Hamming weight $\sum_{i=1}^N x_i =
k$, where each evaluation of $f_\gamma(x)$ requires solving a semidefinite
program—shares key structural similarities with the problem of finding the ground
state of a spin glass.
Both involve searching for a global minimum over a highly combinatorial
configuration space, often riddled with frustration and local minima.
In the spin glass case, the energy landscape is determined by a Hamiltonian,
typically quadratic in spin variables $s_i \in \{-1, +1\}$ and defined explicitly
via couplings and external fields~\cite{mezard1987spin}.
By contrast, in our setting the objective function is not given analytically but
is computed indirectly through the solution of an SDP, which introduces
significant computational overhead.
Nevertheless, the fixed Hamming weight constraint imposes a global restriction
absent in standard spin-glass models, effectively reducing the feasible set to a
lower-dimensional manifold~\cite{lucas2014ising,barahona1982computational}.
Our aim is to render this optimization computationally feasible by developing
strategies that select the most relevant moments for the problem at hand, using
three complementary optimization methods described in Section~\ref{sec:opt_methods}.
The practical difficulty of the moment-selection problem originates in the
inability of $f_\gamma$ to decompose into independent coordinate contributions.
The effect of adding a moment to the relaxation depends on which other moments
are already included. Hence, the best pair of moments need not contain the best single
moment, and the optimal triple is not obtained by extending the optimal pair.
As it will become clearer below, the high-order structure of the solution landscape implies that greedy strategies perform poorly. 

With the optimization problem stated, we now start by a concrete and illustrative benchmark that admits exhaustive verification for adding sets $\mathcal A$ of arbitrary size.

\subsection{Benchmark scenario: the \texorpdfstring{$I_{3322}$}{I3322} Bell inequality}
We apply the optimization framework to the $I_{3322}$ Bell inequality in the
$(3,3,2,2)$ scenario — three measurement settings and two outcomes per party —
one of the simplest Bell inequalities exhibiting nontrivial NPA behavior and a
standard benchmark in the field~\cite{CollinsGisin2004, BrunnerEtAl2014}.
In the correlator representation of~\cite{CollinsGisin2004}, the inequality reads
\begin{equation}
\label{eq:I3322}
I_{3322} =
\langle A_0 B_0 \rangle
+ \langle A_0 B_1 \rangle
+ \langle A_1 B_0 \rangle
+ \langle A_1 B_1 \rangle
- \langle A_0 B_2 \rangle
+ \langle A_1 B_2 \rangle
- \langle A_2 B_0 \rangle
+ \langle A_2 B_1 \rangle
- \langle A_1 \rangle
- \langle B_0 \rangle
- 2\langle B_1 \rangle,
\end{equation}
where $A_i$ ($B_i$) denotes the measurement operator of Alice (Bob) for input $i$
in the Bell test.
The classical bound is $0$, and the best currently known quantum value is
approximately $0.250\,875\,38$, obtained at NPA4 or
higher~\cite{PalVertesi2010}.  Numerical evidence suggests that its true
quantum maximum may require infinite-dimensional quantum
systems~\cite{PalVertesi2010}, making it a challenging and physically
interesting benchmark.  In contrast, the \textsf{NPA1} relaxation gives
$0.375\,000\,01$, and \textsf{NPA2} gives $0.251\,021\,73$.
The adding set $\mathcal{A} =
\textsf{NPA2}\setminus\textsf{NPA1}$ contains $N = 21$ moments, so the full
configuration space $\{0,1\}^{21}$ comprises approximately two million subsets
— small enough for exhaustive enumeration, giving ground-truth solutions
against which all methods are benchmarked.
\subsection{Global landscape of the moment-selection problem}
\begin{figure}[htbp]
    \centering
    \begin{subfigure}[b]{0.48\textwidth}
        \centering
        \includegraphics[width=\textwidth]{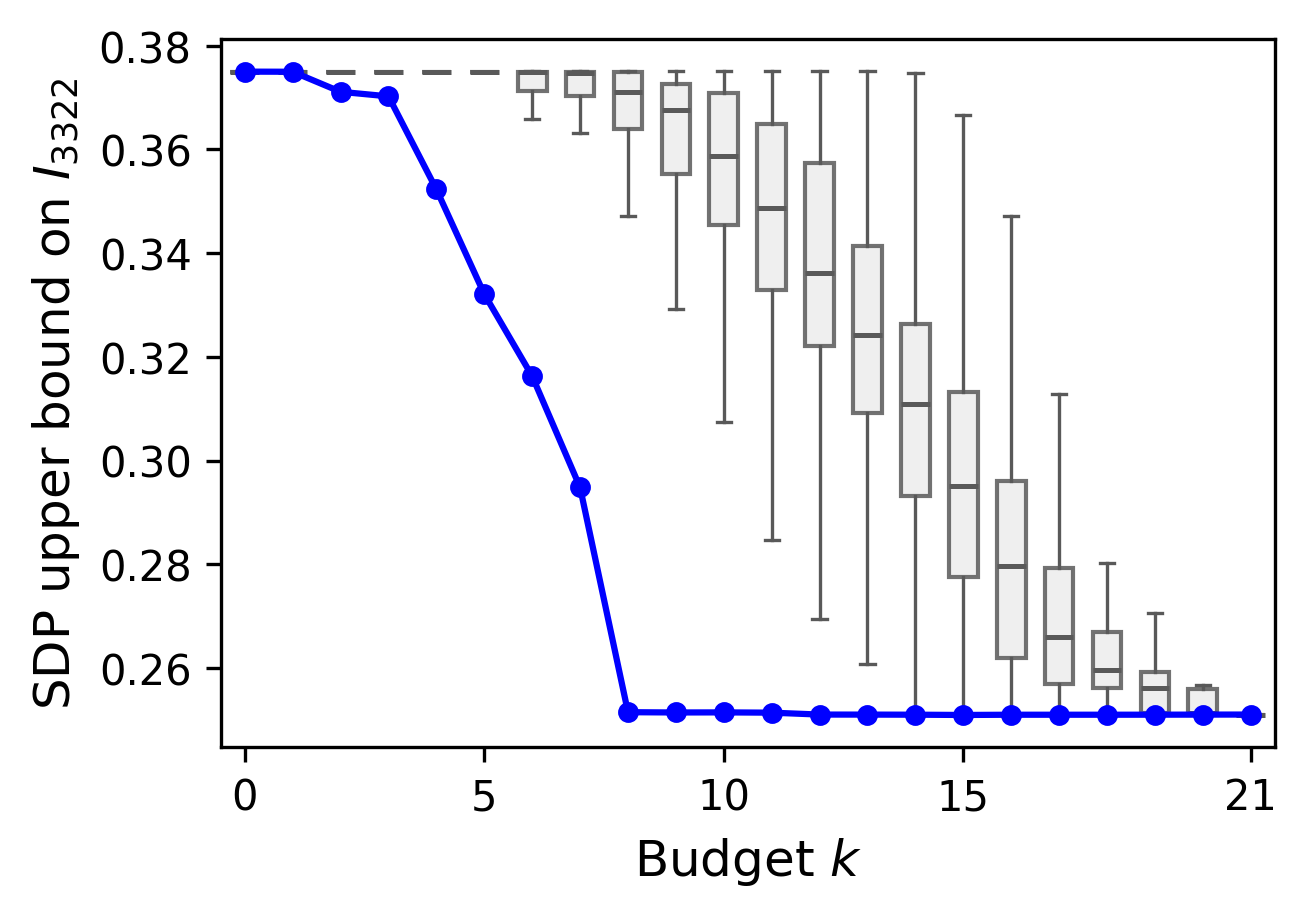}
        \phantomsubcaption\label{fig:boxplot}
    \end{subfigure}
    \hfill
    \begin{subfigure}[b]{0.48\textwidth}
        \centering
        \includegraphics[width=\textwidth]{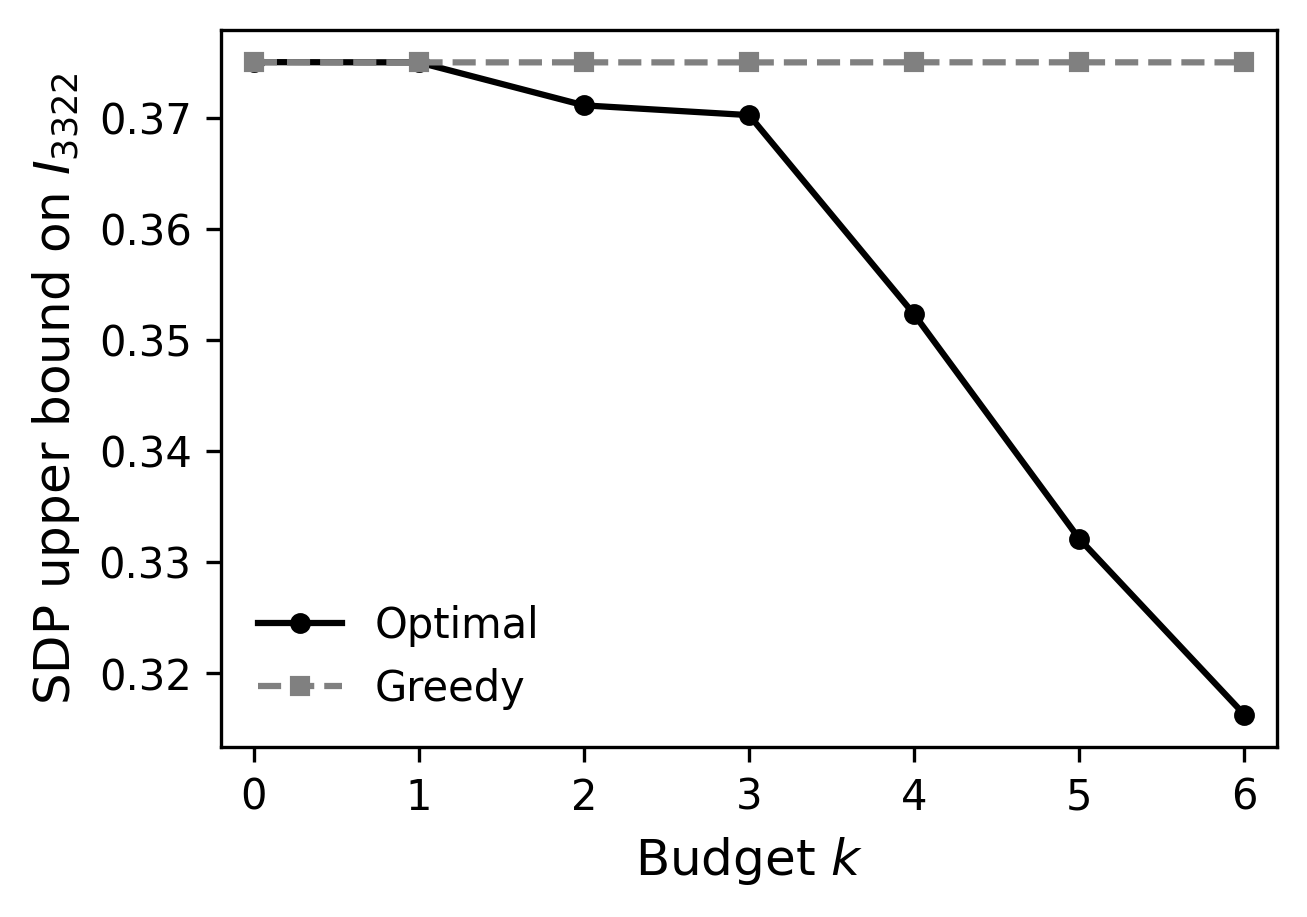}
        \phantomsubcaption\label{fig:density}
    \end{subfigure}
    \caption{\textbf{Complete distribution of $I_{3322}$ relaxation values
    across all $2^{21}$ moment subsets, stratified by Hamming weight $k$.}
    The structural transition in the range $5 < k \leq 19$ — where distributions are
    widest and the optimal value improves most rapidly — identifies the
    combinatorially active regime of the landscape.  The wide spread of bound values at each $k$, with subsets of the same size
    spanning nearly the full range from the \textsf{NPA1} to the \textsf{NPA2}
    bound, is a direct signature of strong higher-order interactions among
    moments.
    \textbf{(a)}~Distribution of relaxation values at each Hamming weight $k$
    over all $\binom{21}{k}$ subsets.  Each box spans the interquartile range
    (Q1--Q3), with the median marked by the interior line and whiskers
    extending to the most extreme values within $1.5\times\mathrm{IQR}$.
    The blue curve is the brute-force optimal (minimum over all subsets of
    size $k$), which reaches the \textsf{NPA2} bound already at $k=8$.
    \textbf{(b)}~Greedy path versus brute-force optimal for $k=0,\ldots,6$.
    The greedy path remains confined to an almost flat plateau near the
    \textsf{NPA1} bound, reflecting the negligible marginal contribution of
    each moment when assessed in isolation.  The optimal value, by contrast,
    drops sharply from $k=4$ onward, revealing that the best subsets cannot
    be obtained by sequential extension and providing direct empirical evidence
    of strong higher-order interactions among moments.}
    \label{fig:combined_distributions}
\end{figure}

The exhaustive enumeration afforded by the $I_{3322}$ benchmark lets us map the full landscape before designing any optimization algorithm.
Figure~\ref{fig:combined_distributions} maps the complete distribution of
$I_{3322}$ relaxation values across all $2^{21}$ moment subsets, stratified
by Hamming weight $k$.  The distribution reveals three distinct regimes.
For $k \leq 5$, the distribution is tightly concentrated near the \textsf{NPA1}
bound of $0.375\,000\,01$; no small subset produces a meaningful improvement,
and the landscape is nearly flat in this region.  This rules out greedy
strategies at the outset, since they would find no useful gradient to follow
and would either terminate prematurely or wander without purpose.
In the transition range $5 < k \leq 19$, the distributions broaden sharply,
developing heavy lower tails that extend all the way to the NPA2 optimum.
The spread between the median and the best achievable value reaches its maximum
in this regime.  This is where the combinatorial structure of the problem is
most pronounced: subsets of the same size can achieve bounds that differ by
nearly the full gap between NPA1 and NPA2.
The landscape is highly heterogeneous in this regime: the typical value across
random subsets is not representative of the best achievable value, and the
landscape exhibits the hallmark features of combinatorial frustration.
For $k > 19$, the distributions concentrate again near the NPA2 value, as
most large subsets already approximate the complete relaxation.  The improvement
in the relaxation bound is therefore not distributed uniformly across $k$:
the vast majority of the gain is concentrated in the transition regime, where
the choice of subset is critical.
\subsection{Synergy as a convergence diagnostic}
\label{subsec:synergy_results}
\begin{figure}[ht]
    \centering
    \begin{subfigure}[b]{0.48\textwidth}
        \centering
        \includegraphics[width=\textwidth]{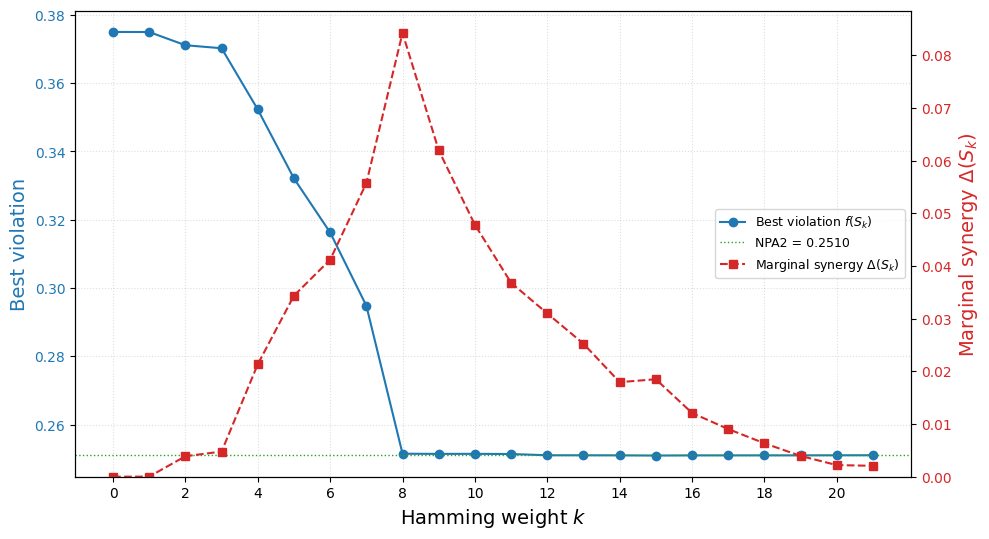}
        \phantomsubcaption\label{fig:synergy-exhaustive}
    \end{subfigure}
    \hfill
    \begin{subfigure}[b]{0.48\textwidth}
        \centering
        \includegraphics[width=\textwidth]{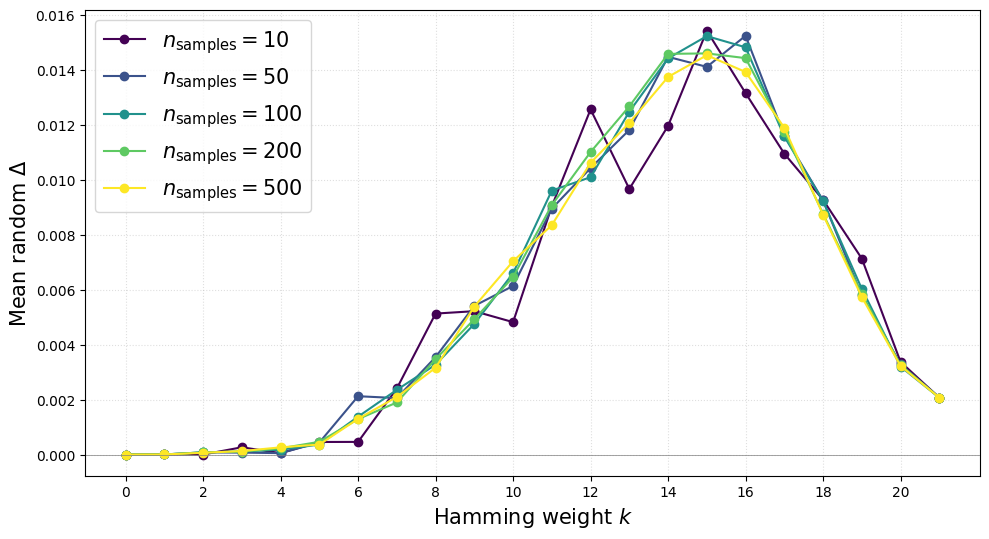}
        \phantomsubcaption\label{fig:synergy-convergence}
    \end{subfigure}
    \caption{\textbf{Synergy as a diagnostic for subset saturation.}
    \emph{Left:} Best violation $f(S_k)$ (blue, left axis) and marginal synergy
    $\Delta(S_k)$ (red, right axis) as a function of $k$, with $S_k$ taken as
    the \emph{globally optimal} $k$-subset of the I3322 NPA1$\to$NPA2 adding
    set. The cost curve flattens abruptly at $k=8$ as the NPA2 bound is reached
    (green dotted line), and the synergy attains its maximum
    $\Delta(S_8^\star) \approx 0.084$ at exactly the same step --- every element
    of the saturating $8$-subset is essential, no single removal preserves
    saturation. Past $k=8$ additional elements are redundant and the synergy
    decays smoothly toward zero.
    \emph{Right:} Convergence of the statistical synergy estimate in
    $n_{\mathrm{samples}}$. Mean random-subset synergy $\overline{\Delta}(k)$
    for five sample budgets, $n_{\mathrm{samples}} \in \{10, 50, 100, 200,
    500\}$, run with a shared seed so that smaller scans are strict subsamples
    of larger ones. The $n=10$ curve is too noisy to locate the peak reliably;
    the bell shape resolves cleanly by $n=50$ and is essentially converged by
    $n=200$. The peak position --- usable as a cheap upper bound on the
    saturation step --- stabilises around $k \approx 14$--$15$ once the sample
    budget is sufficient.}
    \label{fig:synergy}
\end{figure}

The wide variance in the transition regime is not accidental: it reflects a fundamental non-modularity of the landscape, where the contribution of any single moment depends on which others are already included.
To quantify this high-order structure, we adapt the synergy-first backbone
decomposition of~\cite{varley2024synergy}.  Given the best subset $S_k$ found
at budget $k$, the \emph{marginal synergy} is
\begin{equation}
\label{eq:synergy}
    \Delta(S_k)
    := \frac{1}{k}\sum_{i=1}^{k} f_\gamma(S_k \setminus \{m_i\}) - f_\gamma(S_k),
\end{equation}
which measures how much the subsets obtained by removing a single moment
from $S_k$ are \emph{worse} on average than $S_k$ itself.  Because
$f_\gamma$ is being minimized and adding moments tightens the bound,
$f_\gamma(S_k) \leq f_\gamma(S_k \setminus \{m_i\})$ for all $i$, so
$\Delta(S_k) \geq 0$ always.  A large $\Delta(S_k)$ indicates that the
moments in $S_k$ operate collectively; their joint contribution to tightening
the relaxation substantially exceeds the average contribution of any individual
member, so that removing any single moment causes a disproportionate
degradation of the bound.  A value near zero, by contrast, means that
each moment contributes approximately independently.  All evaluations
in~\eqref{eq:synergy} involve subsets of size $k-1$, so the synergy is
computable from the data already gathered at step $k-1$, at no additional
SDP cost when running an exhaustive search.
Figure~\ref{fig:synergy-exhaustive} illustrates this interplay concretely.  The
central feature of the figure is what occurs around $k \approx 9$--$10$: the
cost curve (blue) flattens as the relaxation reaches the \textsf{NPA2} bound,
while the synergy $\Delta(S_k)$ (red) simultaneously peaks and then begins to
decay, jointly marking this as the genuine transition to a modular regime where
saturation has been reached.  In contrast, the earlier plateau of the cost
curve around $k \approx 4$--$6$ is not accompanied by a decay in the synergy,
which remains elevated, correctly signaling that collective effects still
dominate and that further optimization remains worthwhile.
The synergy diagnostic is precisely what distinguishes these two cases.

A complementary and practically more accessible version of the same diagnostic
is obtained by replacing the exhaustive optimal subset with a random sample of
$k$-element subsets drawn uniformly from $\mathcal{A}$.  The mean synergy
$\overline{\Delta}(k)$ over such a sample inherits the same qualitative
bell shape as the exact curve: it rises through the synergy-dominated regime,
peaks near the saturation step, and decays once the landscape becomes modular.
Crucially, the peak of $\overline{\Delta}(k)$ provides a cheap \emph{upper
bound} on the value of $k$ at which the optimizer should be expected to
saturate, and can therefore serve as a budget guide before any dedicated
optimization run is launched.
Figure~\ref{fig:synergy-convergence} shows that this peak is already
locatable with moderate sample budgets: the bell shape is well resolved by
$n_{\mathrm{samples}} = 50$ and the peak position stabilises around
$k \approx 15$--$16$ from $n = 200$ onward, with the five curves in close
agreement at all $k$ except the very noisiest ($n=10$) run.  The consistency
across sample sizes also provides a practical convergence check: once the peak
location stops shifting as $n_{\mathrm{samples}}$ is increased, the estimate
can be trusted.
Notably, the synergy diagnostic is computed entirely from subsets of size $k-1$
and therefore requires no additional SDP evaluations beyond those already
performed during an exhaustive search, making it a computationally cost-free
supplement to the standard convergence monitor.

\section{Optimization methods}
\label{sec:opt_methods}

In the previous section, a brute-force approach to optimally select moments for the $I_{3322}$ Bell inequality is possible because the space of possible choices is manageable. This is however not possible when considering other, more complex, noncommutative polynomial optimisations. To navigate the landscape of possibilities we consider three different optimisation methods, which search for good moment selections with no guarantee of optimality. The details of these three methods are described in this section. Readers primarily interested in their performance may skip what follows and proceed directly to Section~\ref{sec:results}, where they are applied to other Bell inequalities and ground-state problems.

\subsection{Optimization method 1: Monte Carlo approaches}
\label{subsec:MC}
Monte-Carlo approaches to moment selection will select candidate moments $x_t\in S_N^k$ via rejection sampling. Let us begin with Simulated Annealing (SA)~\cite{kirkpatrick1983optimization}. SA explores the
landscape via the Metropolis--Hastings criterion~\cite{metropolis1953equation}.
Starting from $x_t \in S_N^k$, a candidate $x'$ is generated by a random
Hamming-weight-preserving bit flip and accepted with probability
\begin{equation}
A(x_t \rightarrow x') = \min\!\left(1,
\exp\!\left(-\frac{f_\gamma(x') - f_\gamma(x_t)}{T}\right)\right),
\end{equation}
with temperature decaying as $T(t) = T_0 \cdot \alpha^t$.  While SA provides
a natural starting point by exploring local neighborhoods
of the
configuration space, its performance may degrade in the presence of rugged
energy landscapes with high barriers separating competing minima.
To further
enhance global exploration and enable parallelization, we employ
\emph{Parallel Tempering} (PT), also known as \emph{Replica Exchange Monte Carlo}
\cite{hukushima1996exchange,earl2005parallel}.  Parallel Tempering generalizes
the simulated annealing paradigm by evolving multiple replicas of the system
in parallel at different temperatures and allowing configuration exchanges
between them.
In PT, we consider a collection of $R$ replicas, each associated with a fixed
temperature $T_1 < T_2 < \cdots < T_R$.  Each replica independently performs a
local stochastic optimization—here implemented as a short simulated annealing
run at constant temperature—sampling from the Boltzmann distribution
\begin{equation}
\pi_{T_r}(x) \propto \exp\!\left(-\frac{f_\gamma(x)}{T_r}\right),
\end{equation}
where $f_\gamma(x)$ is the cost function defined in
Eq.~\eqref{eq:opt_problem}.  Low-temperature replicas focus on exploitation near
low-cost configurations, while high-temperature replicas explore the landscape
more broadly and can cross large energy barriers.
At regular intervals, exchange moves between pairs of replicas at neighboring
temperatures are proposed.  Given two replicas at temperatures $T_i$ and
$T_j$, currently in configurations $x_i$ and $x_j$ with costs
$f_\gamma(x_i)$ and $f_\gamma(x_j)$, the swap is accepted with probability
\begin{equation}
\label{eq:pt_swap}
A_{\text{swap}} =
\min\!\left(
1,
\exp\!\left[
\left(\frac{1}{T_i} - \frac{1}{T_j}\right)
\left(f_\gamma(x_j) - f_\gamma(x_i)\right)
\right]
\right),
\end{equation}
which ensures detailed balance with respect to the joint distribution
$\prod_r \pi_{T_r}(x_r)$ \cite{hukushima1996exchange}.
We emphasize that replica exchanges are performed using the \emph{current}
configurations of each chain, rather than their historically best states, in
order to preserve detailed balance and the Markovian structure of the
dynamics~\cite{levin2017markov}.
The best configuration encountered across all replicas is tracked separately
for optimization purposes only.
In our implementation, each replica executes a fixed number of simulated
annealing steps at constant temperature before exchange attempts are made.
This structure allows the computationally expensive evaluations of
$f_\gamma(x)$—which require solving a semidefinite program—to be naturally
parallelized across replicas.  Parallel Tempering is particularly well suited
to the present setting, as it combines efficient barrier crossing with
parallel execution, while preserving the fixed Hamming-weight constraint
$x \in S_N^k$ in all local updates.
\begin{algorithm}[H]
\caption{Parallel Tempering for moment subset optimization}
\label{alg:PT}
\begin{algorithmic}[1]
\Require Cost function $f_\gamma(x)$, $R$ replicas, geometric temperature
ladder $T_1 > \cdots > T_R$, steps per epoch $L$, number of epochs $E$
\State Initialize replicas $\{x_r^{(0)} \in S_N^k\}_{r=1}^R$ uniformly
\State Evaluate $f_\gamma(x_r^{(0)})$ for all $r$
\State $x_{\text{best}} \gets \arg\min_r f_\gamma(x_r^{(0)})$
\For{$e = 1$ to $E$}
    \ForAll{replicas $r = 1,\dots,R$ \textbf{in parallel}}
        \State Run $L$ Metropolis steps at fixed $T_r$
        \State Update $x_r$ and track $x_{\text{best}}$
    \EndFor
    \ForAll{pairs $(r,\, r+1)$ with $r \equiv e \pmod{2}$}
        \State Propose swap; accept with probability $A_{\text{swap}}$
        (Eq.~\eqref{eq:pt_swap})
    \EndFor
\EndFor
\State \Return $x_{\text{best}},\ f_\gamma(x_{\text{best}})$
\end{algorithmic}
\end{algorithm}
\subsection{Optimization method 2: Restricted Boltzmann Machine}
At a high level, our goal is to choose a binary vector $x \in S^k_N:=\{x \in \{0,1\}^N \; s.t. \; ||x||_1 = k\}$ which specifies a subset of moments to include in a relaxation. Each choice induces a concrete SDP instance whose solution yields a scalar objective value (i.e the violation),
\begin{equation}
    \mathcal{L}(x) = f_{\gamma}(x).
\end{equation}
From an optimization perspective, $f_{\gamma}$ is a black-box; its output is obtained by running an SDP solver, and there is no tractable path to differentiating $\mathcal{L}(x)$ with respect to the internal parameters of a sampling model that generated $x$. Moreover, the space of possible actions is combinatorial, and thus the sampling step is itself discrete.
These factors naturally suggest an online reinforcement-learning (RL) viewpoint for solving this problem \cite{williams1992simple, barto2021reinforcement}. We are interacting with an environment that, given some action $x \in S^k_N$, returns a scalar feedback $\mathcal{L}(x)$. We can solve for optimal actions by optimizing a parametric policy $\pi_{\theta}(x)$ that concentrates probability on low-loss subsets. We note here that this formulation is an \textit{online} RL problem that runs per-instance rather than offline. Each new optimization instance defines an independent problem with a different black-box landscape over $S^k_N$, and hence in general requires learning a new policy from scratch.
There is no fixed dataset of optimal actions, and data is generated on-the-fly by querying the SDP solver during training.
Hence, let us consider the following policy optimization objective,
\begin{equation}
    J(\theta) := \mathbb{E}_{x \sim \pi_{\theta}}\left[ \mathcal{L}(x)\right],
\end{equation}
where we seek $\theta^* = \operatorname{argmin}_{\theta} J(\theta)$. 

A natural RL algorithm to employ in this setting is REINFORCE \cite{barto2021reinforcement}. This is for several reasons. First, $\mathcal{L}$ is a black-box and $x$ is discrete, which is compatible with REINFORCE. Because the interaction is effectively a one-step contextual bandit, there is no temporal credit assignment to exploit. That is, there is no aggregation of the environment signal or a time-horizon for it. Hence, value-based bootstrapping and actor-critic methods offer little leverage here, and  mainly act as learned baselines. They could be useful in principle, but they contain unnecessary complexity given our one-step-horizon setting.
Let $\pi_{\theta}(x)$ be a parametric distribution over $x \in S^k_N$, with variational parameters $\theta \in \mathbb{R}^n$. The policy gradient theorem allows us to perform gradient based updates to our policy via the objective function,
\begin{equation}
    \nabla_{\theta} J(\theta) = \mathbb{E}_{x \sim \pi_{\theta}}\left[\mathcal{L}(x) \nabla_{\theta} \log \pi_{\theta}(x)\right].
\end{equation}
This enables updates because crucially, we never have to differentiate through the environment, only the sampler itself (i.e. the policy). Intuitively, the factor $\nabla_{\theta} \log \pi_{\theta}(x)$ points in the direction that would increase the probability of the sampled action $x$, and the scalar $\mathcal{L} (x)$ weights the importance of this direction by the quality of the yielded SDP solution.  Hence, in gradient descent on $J$, samples with lower-than-typical cost are reinforced (i.e. their log-probability increases), while high-cost samples are suppressed.
To reduce variance, we may subtract any baseline, $b\in\mathbb{R}$ that is independent of the current sampled action without changing the expectation,
\begin{equation}
    \nabla_{\theta} J(\theta) = \mathbb{E}_{x \sim \pi_{\theta}}\left[(\mathcal{L}(x) - b) \nabla_{\theta} \log \pi_{\theta}(x)\right].
\end{equation}
Among RL practitioners, a common choice is the exponential moving average (EMA) of observed losses
\begin{equation}
    b \gets \beta b + (1- \beta) \mathcal{L}(x),
\end{equation}
where $\beta \in (0,1]$. Employing EMA averages means $\mathcal{L}(x) - b$ acts as an advantage-like signal. Samples better than the EMA average increase their probability, and worse-than-typical samples are suppressed.

To implement REINFORCE with standard autodiff, we can define our loss function as
\begin{equation}
    \tilde{\mathcal{L}}_{\theta}(x) = -(\mathcal{L}(x) - b) \log \pi_{\theta}(x), \label{eq:REINFORCE_loss}
\end{equation}
since $\nabla_{\theta}\tilde{\mathcal{L}}_{\theta}$ is the stochastic descent direction for $J(\theta)$.
Let us now specify the form of the policy $\pi_{\theta}$. We use a \textit{Restricted Boltzmann Machine (RBM)}, which serves as a parametric distribution over binary vectors \( x \in S_N^k \). In standard RBM sampling, one draws a hidden vector \( h \) from the posterior $p(x | h)$, then samples a new visible vector \( x \) from the conditional distribution $p(x | h)$, typically using sigmoid activation followed by Bernoulli sampling \cite{hinton2002training, hinton2006fast, hinton2012practical, goodfellow2016deep, heightman2025deep}.
However, in this scheme we are presented with a further constraint. Namely,  we aim to sample only those vectors that have fixed Hamming weight, i.e. \( \lVert x \rVert_1 = k \). This is because this binary vector corresponds to a choice of which moments to include, so its Hamming weight is constrained by the amount of moments to be included. To enforce this constraint, we use the Gumbel trick \cite{tucker2017rebar, titsias2015hamming, maddison2016concrete}, which is detailed below.
Let the hidden vector \( h \in \{0,1\}^{*} \) be sampled from
\(h \sim p(h \mid x) = \sigma(Wx + b^h),\), where \( W \) is the RBM weight matrix, \( b^h \) is the hidden bias vector, and \( \sigma(z) = \frac{1}{1 + e^{-z}} \) is the element-wise sigmoid function. We may then compute the visible-layer logits in the usual way, \(y_i = (W^\top h)_i + b^v_i,\), with corresponding activation probabilities,
\[
    p_i := \sigma(y_i) = \frac{1}{1 + e^{-((W^\top h)_i + b^v_i)}}.
\]
To sample a binary vector \( x \in S_N^k \) with fixed Hamming weight \( k \), we introduce Gumbel noise \cite{huijben2022review, bertin2005global} \( g_i \sim \mathrm{Gumbel}(0,1) \) and define perturbed scores,
\begin{equation}
\label{eq:gubel_reparam}
    s_i := \log p_i + g_i.
\end{equation}
We then define the sampled vector \( x^k \in S_N^k \) by selecting the indices of the highest $k$ values of the perturbed scores (i.e. top-\( k \) ) and setting those positions to 1,
\[
    x^k_i =
    \begin{cases}
        1 & \text{if } i \in \operatorname{Top-k}(s, k), \\
        0 & \text{otherwise}.
    \end{cases}
\]
The top-$k$ operation produces valid discrete actions $x^k \in S^k_N$. Each such sample \( x^k \) defines a moment relaxation, for which we solve the corresponding SDP and obtain a scalar loss,
\[
    \mathcal{L}(x^k) := f_\gamma(x^k).
\]
We emphasize here that in our implementation, the policy $\pi_\theta$ is defined operationally as the distribution over $x^k \in S_N^k$ obtained by running $M$ alternating RBM updates (sampling $h \sim p_\theta(h\mid x)$, then forming $p_\theta(x\mid h)$) with a top-$k$ projection at the visible layer. This is distinct from the typical setting in which the RBM's energy is employed as a stationary distribution over binary vectors.
Since this map is non-differentiable and $\mathcal{L}$ is a black-box, we can optimize $\theta$ via REINFORCE rather than differentiating pathwise via the reparameterization trick. The RBM's variational parameters, \( \theta = \{W, b^v, b^h\} \), instead get updates via the policy gradient theorem detailed above. We note here that REINFORCE requires $\log \pi_{\theta}$, which we can estimate with a proxy derived from the RBM's visible conditionals. Treating $p_i = \sigma(y_i)$ as Bernoulli probabilities, we may write
\begin{equation}
    \log \pi_{\theta} \approx \sum_{i = 1}^N \left[x^k_i \log p_i + (1 - x^{k}_i) \log (1 - p_i)\right]. \label{eq:proxy_log_policy}
\end{equation}
This is the conditional Bernoulli log-likelihood under $p(x | h)$, and is not the exact probability of a top-$k$ sample. While the sampling procedure directly enforces the top-$k$ constraint, this likelihood proxy provides a score function direction that increases the logits of indices frequently present in low-loss subsets and decreases the rest. Empirically, we observed that this is sufficient to bias exploration towards useful moment sets under a fixed evaluation budget.
We now proceed to detail the initialization, architecture and hyperparameters to train our model.
The weight matrix in the RBM was initialized with a random Normal $W_{ij} \sim \mathcal{N}(0,0.01)$ whilst the biases on each node were initialized to $0$. To ensure an architecture which scales well with the adding set, the number of hidden nodes in the RBM is set to \texttt{int}$(|S_N|/2)$, meaning the total number of nodes in the architecture is $N + N /2$ which is $\mathcal{O}(N)$.
The number of training steps of the RBM was set to 500, with 10 MCMC steps per train-step. For the optimizer, we opt for ADAM with a learning rate of $0.01$. Finally, we hold a local variable \texttt{current\_vec} which is the optimal $k$-hot vector encountered during training, with its corresponding \texttt{current\_cost} holding the optimal violation corresponding to \texttt{current\_vec}. The reason for this is that unlike typical use-cases for an RBM, there is no deployment phase. We are simply using it as a parametric distribution. As such there is no convergence per-se, rather we hold the number of training steps to match the run-time of the other two methods detailed above. A summary of this method is provided in pseudocode below.
\begin{algorithm}[H]
\caption{REINFORCE with RBM policy}
\label{alg:rbm_reinforce}
\begin{algorithmic}[1]
\Require Initial subset $x^k \in S_N^k$, black-box loss $\mathcal{L}(x)=f_\gamma(x)$ (via SDP solve), RBM parameters $\theta=\{W,b^v,b^h\}$,
number of policy updates $T$, Gibbs/MCMC steps per update $M$, Hamming weight $k$,
learning rate $\eta$, baseline decay $\beta \in (0,1]$, (optional) cooling schedule $\{\tau_t\}_{t=1}^T$
\State Initialize $W_{ij}\sim \mathcal{N}(0,0.01)$, $b^v \gets 0$, $b^h \gets 0$
\State Initialize baseline $b \gets 0$
\State Evaluate $\ell \gets \mathcal{L}(x^k)$; set $x^{\mathrm{best}}\gets x^k$, $\ell^{\mathrm{best}}\gets \ell$
\For{$t=1,\dots,T$}
    \State $x \gets x^k$ \Comment{start from current subset}
    \For{$m=1,\dots,M$}
        \State Sample hidden units: $h \sim \mathrm{Bernoulli}\!\left(\sigma(Wx+b^h)\right)$
        \State Compute visible logits: $y \gets W^\top h + b^v$; probabilities $p \gets \sigma(y)$
        \State Sample Gumbel noise: $g_i \sim \mathrm{Gumbel}(0,1)$ for $i=1,\dots,N$
        \State Compute scores: $s_i \gets (\log p_i + g_i)/\tau_t$ \Comment{set $\tau_t \equiv 1$ if no cooling}
        \State Project to $S_N^k$: $x \gets \operatorname{Top-k}(s,k)$
    \EndFor
    \State Set $x^k \gets x$ \Comment{new action / subset}
    \State Evaluate black-box loss: $\ell \gets \mathcal{L}(x^k)$ \Comment{solve SDP}
    \State Update baseline (EMA): $b \gets \beta b + (1-\beta)\ell$
    \State Compute proxy log-likelihood, $\log \pi_\theta(x^k)$, via Eq.~(\ref{eq:proxy_log_policy}).
    \State Compute REINFORCE loss $\tilde{\mathcal{L}}_\theta$ via Eq.~(\ref{eq:REINFORCE_loss}).
    \State Update parameters: $\theta \gets \mathrm{Adam}(\theta, \nabla_\theta \tilde{\mathcal{L}}_\theta, \eta)$
    \If{$\ell < \ell^{\mathrm{best}}$}
        \State $x^{\mathrm{best}} \gets x^k$, \quad $\ell^{\mathrm{best}} \gets \ell$
    \EndIf
\EndFor
\State \Return $x^{\mathrm{best}}, \ell^{\mathrm{best}}$
\end{algorithmic}
\end{algorithm}
\subsection{Optimization method 3: Bayesian Optimization}
\label{subsec:BO}
As a third method to solve Eq.~\ref{eq:opt_problem}, we explore Bayesian
Optimization (BO)~\cite{shahriari2016taking}.
BO is designed for objective functions that are expensive to evaluate, possibly
noisy, and non-differentiable: precisely the setting we can face for some instances, where each
evaluation of $f_\gamma$ may require solving a very large semidefinite program.
The method maintains a probabilistic surrogate model of $f_\gamma$ that is cheap
to query, and uses an acquisition function to determine which configuration to
evaluate next, thereby reducing the total number of expensive SDP calls needed
to locate a near-optimal solution.

We adopt a random forest as the surrogate model~\cite{breiman2001random},
following the spirit of sequential model-based
optimization~\cite{hutter2011smac}.
Random forests are well suited to this setting: they are robust to randomly
sampled training data, handle discrete inputs naturally, and provide both a
mean prediction $\mu(x)$ and an uncertainty estimate $\sigma(x)$ via the
variance across the ensemble of regression trees.
For the acquisition function we use the Upper Confidence Bound (UCB) criterion,
which selects the next configuration by balancing exploitation of predicted
low-cost regions against exploration of poorly sampled areas:
\begin{equation}
    x^*=\underset{x\in \{x_c\}}{\arg\min}\ \text{UCB}(x),
    \qquad
    \text{UCB}(x):=\mu(x) - \beta \sigma(x),
\end{equation}
where $\beta>0$ controls the exploration–exploitation trade-off.

The algorithm proceeds as follows.
An initial set of $n_{\text{init}}$ configurations is drawn uniformly at random
from $S_N^k$ and evaluated exactly via SDP, providing the training data for a
first fit of the random forest.
In the main loop, at each of $n_{\text{iter}}$ iterations a pool of $n_c$
candidate configurations is sampled uniformly from $S_N^k$; the surrogate
provides a prediction $\{\mu(x), \sigma(x)\}$ for each candidate without any
SDP evaluation, and the candidate minimizing the UCB criterion is selected for
exact evaluation.
The training set is extended with this new observation and the surrogate is
refitted before the next iteration.
After all iterations, the configuration with the lowest observed $f_\gamma$
value among the $n_{\text{init}} + n_{\text{iter}}$ evaluations is returned as
the final solution.

The method requires specifying, beyond the shared parameters of the
optimization problem, three additional hyperparameters: $n_{\text{init}}$, the
size of the initial random training set; $n_{\text{iter}}$, the number of BO
iterations; and $n_c$, the number of candidate configurations drawn at each
acquisition step.
The trade-off parameter $\beta$ is held fixed throughout but could in principle
be annealed for improved performance.
A pseudocode summary is given in Algorithm~\ref{alg:BO}.
\begin{algorithm}[H]
\caption{Bayesian Optimization with Random Forest surrogate}
\label{alg:BO}
\begin{algorithmic}[1]
\Require $f_\gamma$, $N$, $k$, $n_{\mathrm{init}}$, $n_{\mathrm{iter}}$,
$n_c$, $\beta$
\State Sample $n_{\mathrm{init}}$ configurations uniformly from $S_N^k$,
evaluate $f_\gamma$, form $\mathcal{D}$
\State Fit RF on $\mathcal{D}$
\For{$t = 1$ to $n_{\mathrm{iter}}$}
    \State Sample candidate set $\mathcal{C} \subset S_N^k$
    \State $x^* \gets \arg\min_{x \in \mathcal{C}}
           [\mu_{\mathrm{RF}}(x) - \beta\,\sigma_{\mathrm{RF}}(x)]$
    \State Evaluate $f_\gamma(x^*)$, add to $\mathcal{D}$, refit RF
\EndFor
\State \Return configuration in $\mathcal{D}$ with minimum $f_\gamma$
\end{algorithmic}
\end{algorithm}

\section{Results}
\label{sec:results}
\subsection{Optimization methods' results and discussion}
\label{subsec:discussion}

As a first comparison, we come back to the $I_{3322}$ Bell inequality under first and second NPA levels, where we have a complete understanding of the moment selection problem. All results in this section were obtained with the following fixed
hyperparameter configuration.  
\begin{itemize}
    \item PT used $R = 5$ chains on a geometric
temperature ladder $T \in [0.1, 2.0]$, for $E = 20$ epochs of $L = 50$
Metropolis steps each, giving $5\,000$ SDP evaluations per run.  
    \item The RBM was
trained for $500$ gradient steps using ADAM ($\eta = 0.01$, decay $= 0.98$),
with $n_{\mathrm{mcmc}} = 10$ internal MCMC steps per gradient update and an
annealing schedule $T_{\mathrm{start}} = 1.0$, $\alpha = 0.99$, giving $501$
SDP evaluations per run (including one initial evaluation).  
    \item BO used
$n_{\mathrm{init}} = 50$ random initialization points, $n_{\mathrm{iter}} =
200$ iterations, $\beta = 0.7$, $n_c = 100$ candidates per iteration, and a
random forest of $100$ trees, giving $250$ SDP evaluations per run. 
\end{itemize} 
All
methods were run for $30$ independent experiments at each value of $k \in \{0,
1, \ldots, 21\}$.
Two reference curves appear throughout: the solid black line is the
ground-truth optimal path from exhaustive enumeration over all $2^{21}$
subsets, and the dashed black line is the greedy path.  For reference, the
brute-force cost at the hardest point $k=10$ is $\binom{21}{10} = 352\,716$
SDP evaluations.  PT, RBM, and BO therefore operate at factors of
approximately $70$, $704$, and $1\,411$ below brute force, respectively,
confirming that all three methods remain well within a tractable computational
regime.

\begin{figure}[htbp]
    \centering
    \begin{subfigure}[b]{0.32\textwidth}
        \centering
        \includegraphics[width=\textwidth]{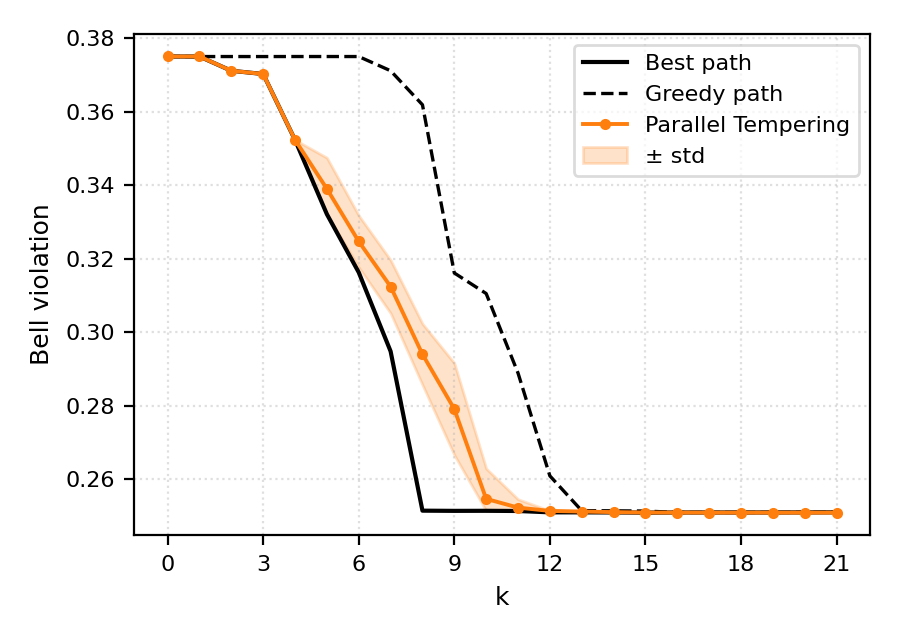}
        \phantomsubcaption\label{fig:pt_new}
    \end{subfigure}
    \hfill
    \begin{subfigure}[b]{0.32\textwidth}
        \centering
        \includegraphics[width=\textwidth]{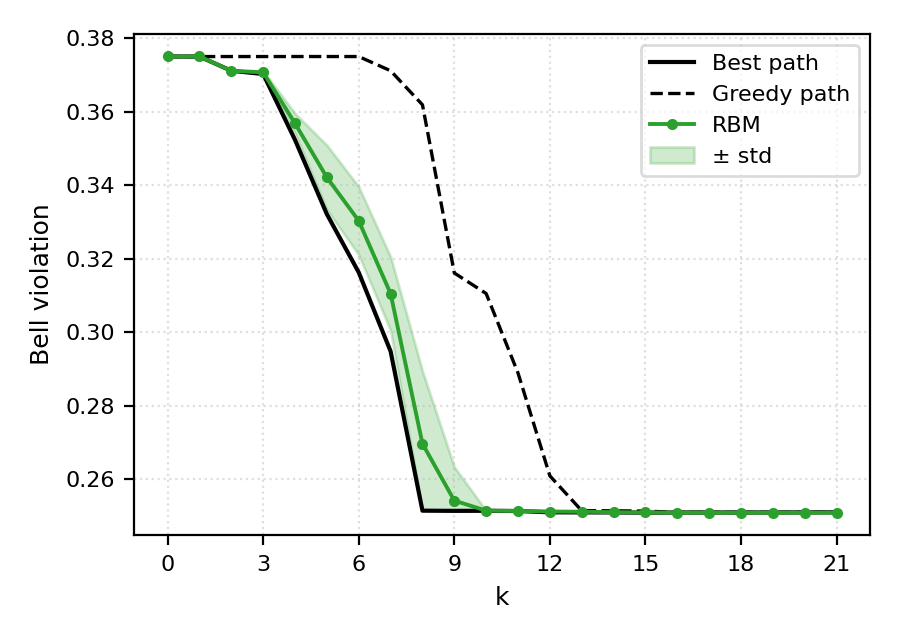}
        \phantomsubcaption\label{fig:rbm_new}
    \end{subfigure}
    \hfill
    \begin{subfigure}[b]{0.32\textwidth}
        \centering
        \includegraphics[width=\textwidth]{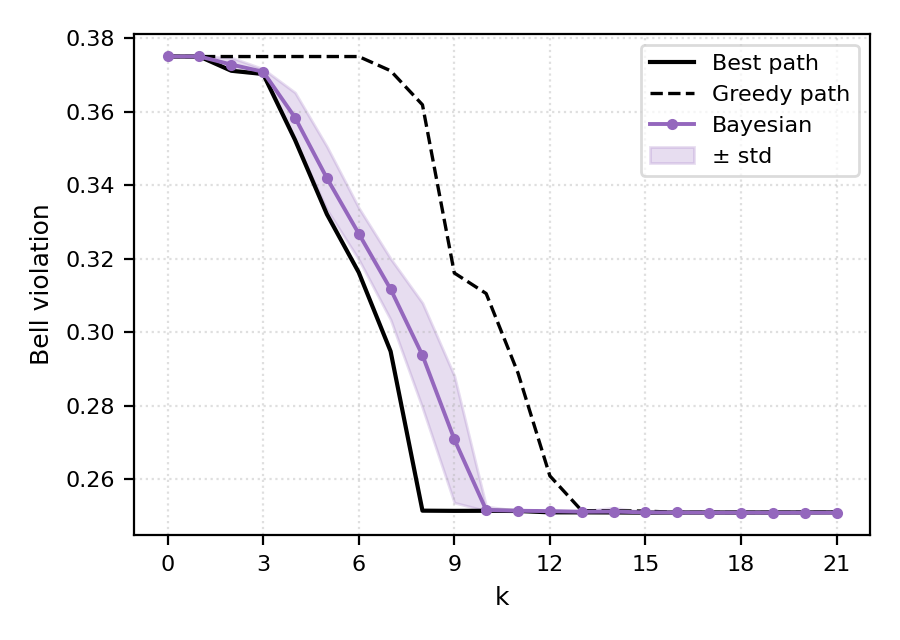}
        \phantomsubcaption\label{fig:bayes_new}
    \end{subfigure}
    \caption{\textbf{Individual method performance on the $I_{3322}$
    moment-selection problem.}  Each panel shows the mean $I_{3322}$ violation
    (colored curve with markers) as a function of $k$, averaged over $30$
    independent runs, with the shaded band indicating one standard deviation.
    The solid black curve is the ground-truth optimal path; the dashed black
    curve is the greedy baseline.  
    \textbf{(a)} PT ($R=5$, $T \in [0.1,2.0]$, $5\,000$ SDP evaluations)
    tracks the optimal curve closely from $k \approx 3$, with a moderate
    variance band that narrows as $k$ passes the transition region.
    \textbf{(b)} RBM ($500$ gradient steps, ADAM $\eta=0.01$, $501$ SDP
    evaluations) achieves close tracking of the optimal path, reaching the
    \textsf{NPA2} convergence value at $k \approx 9$--$10$ with moderate
    variance, demonstrating that gradient-based learning of the sampling
    distribution captures the collective structure of the landscape.
    \textbf{(c)} BO ($250$ SDP evaluations total) exhibits a systematic gap
    relative to the optimal path in the transition regime, with a wider
    uncertainty band, reflecting the difficulty of surrogate modeling under
    strong higher-order interactions.  All three methods substantially
    outperform the greedy baseline, which begins to improve only at
    $k \approx 8$ and does not reach the \textsf{NPA2} bound until
    $k \approx 12$.}
    \label{fig:individual_methods}
\end{figure}

Figure~\ref{fig:individual_methods} shows the performance of each method
with uncertainty bands.
PT tracks the optimal curve closely from $k \approx 3$, with a moderate
variance band that narrows beyond the transition region.  The replica-exchange
mechanism navigates the high-barrier structure of the synergy-dominated regime
effectively, enabling consistent descent toward the NPA2 optimum well before
the greedy path begins to improve.
The RBM achieves the closest tracking of the optimal path, reaching the convergence
value at $k \approx 9$--$10$ with moderate variance.  With $501$ SDP calls,
its gradient-based adaptation of the sampling distribution allows it to learn
and exploit the collective landscape structure in a way that memoryless
Metropolis and the random forest surrogate sampling cannot.
All methods converge to the \textsf{NPA2} value for
sufficiently large $k$.

\begin{figure}[htbp]
    \centering
    \begin{subfigure}[b]{0.48\textwidth}
        \centering
        \includegraphics[width=\textwidth]{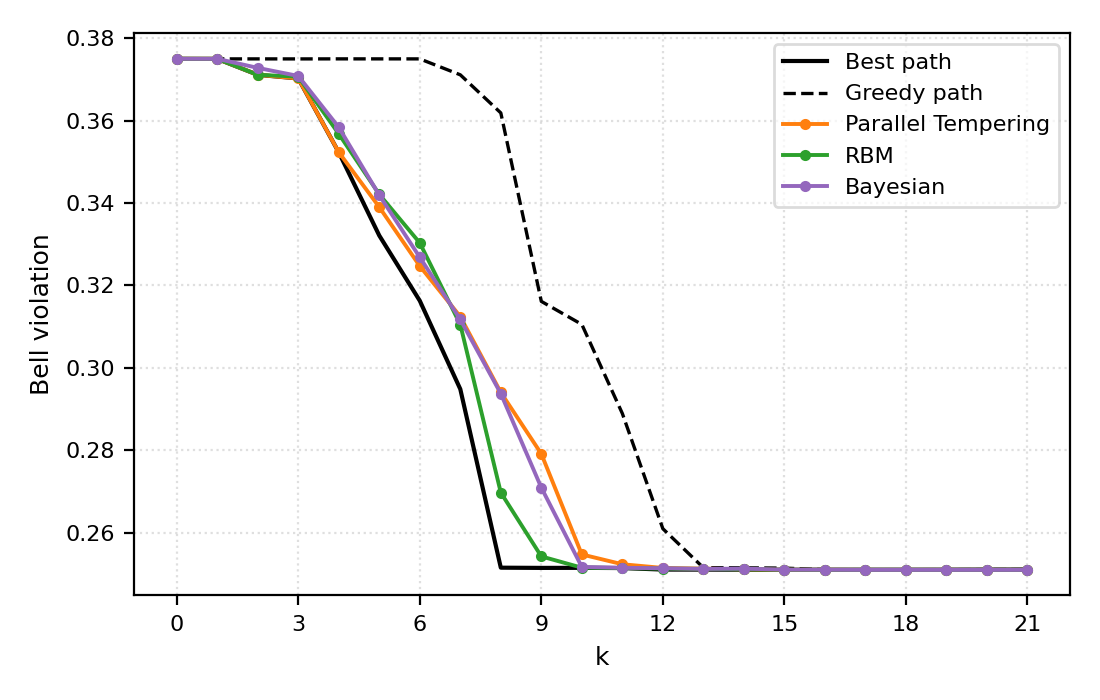}
        \phantomsubcaption\label{fig:comparison}
    \end{subfigure}
    \hfill
    \begin{subfigure}[b]{0.48\textwidth}
        \centering
        \includegraphics[width=\textwidth]{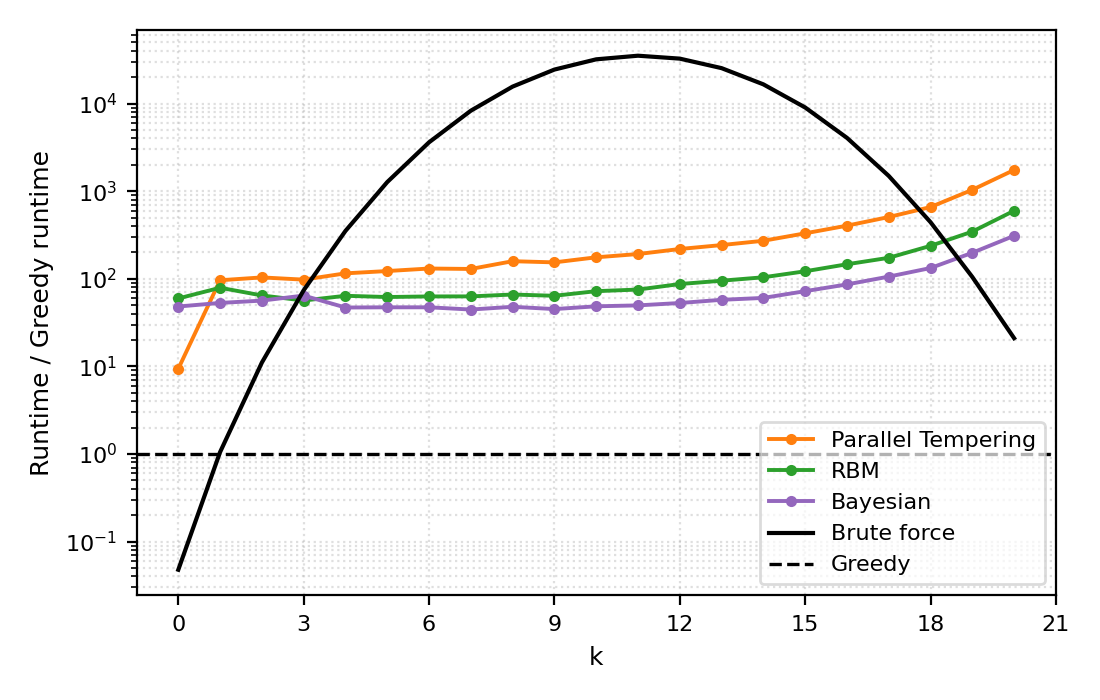}
        \phantomsubcaption\label{fig:runtime}
    \end{subfigure}
    \caption{\textbf{Solution quality and computational cost of the three
    optimization methods on the $I_{3322}$ moment-selection problem.}
    \textbf{(a)}~Mean $I_{3322}$ violation achieved by PT (orange), RBM
    (green), and BO (purple) as a function of~$k$, averaged over $30$
    independent runs.  The solid black curve is the ground-truth optimal path
    from exhaustive enumeration; the dashed black curve is the greedy baseline.
    PT tracks the optimum closely from $k \approx 3$; the RBM achieves a
    substantially closer approach to optimal throughout the hard transition regime
    despite ten times fewer SDP evaluations (see Fig.~\ref{fig:log_distance}).
    BO converges later
    but requires a factor of~$20$ fewer SDP evaluations than PT.
    \textbf{(b)}~Runtime of each method normalized by the greedy cost, on a
    logarithmic scale.  The solid black curve shows the brute-force cost
    $\binom{21}{k}$ under the same normalization, peaking at $352\,716$ at
    $k=10$.  BO remains the least expensive method throughout, within one to
    two orders of magnitude of the greedy baseline; the RBM incurs moderate
    overhead; PT is the most expensive due to its five-chain structure, yet
    stays orders of magnitude below brute force.  Comparing the two panels,
    the RBM offers the best trade-off between accuracy and cost, while BO is
    preferable when SDP evaluations dominate the budget.}
    \label{fig:comparison_runtime}
\end{figure}

\begin{figure}[htbp]
    \centering
    \begin{subfigure}[b]{0.48\textwidth}
        \centering
        \includegraphics[width=\textwidth]{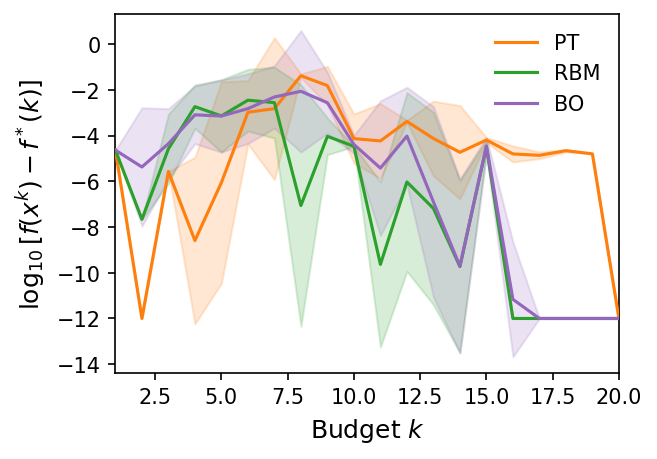}
        \phantomsubcaption\label{fig:log_distance_log}
    \end{subfigure}
    \hfill
    \begin{subfigure}[b]{0.48\textwidth}
        \centering
        \includegraphics[width=\textwidth]{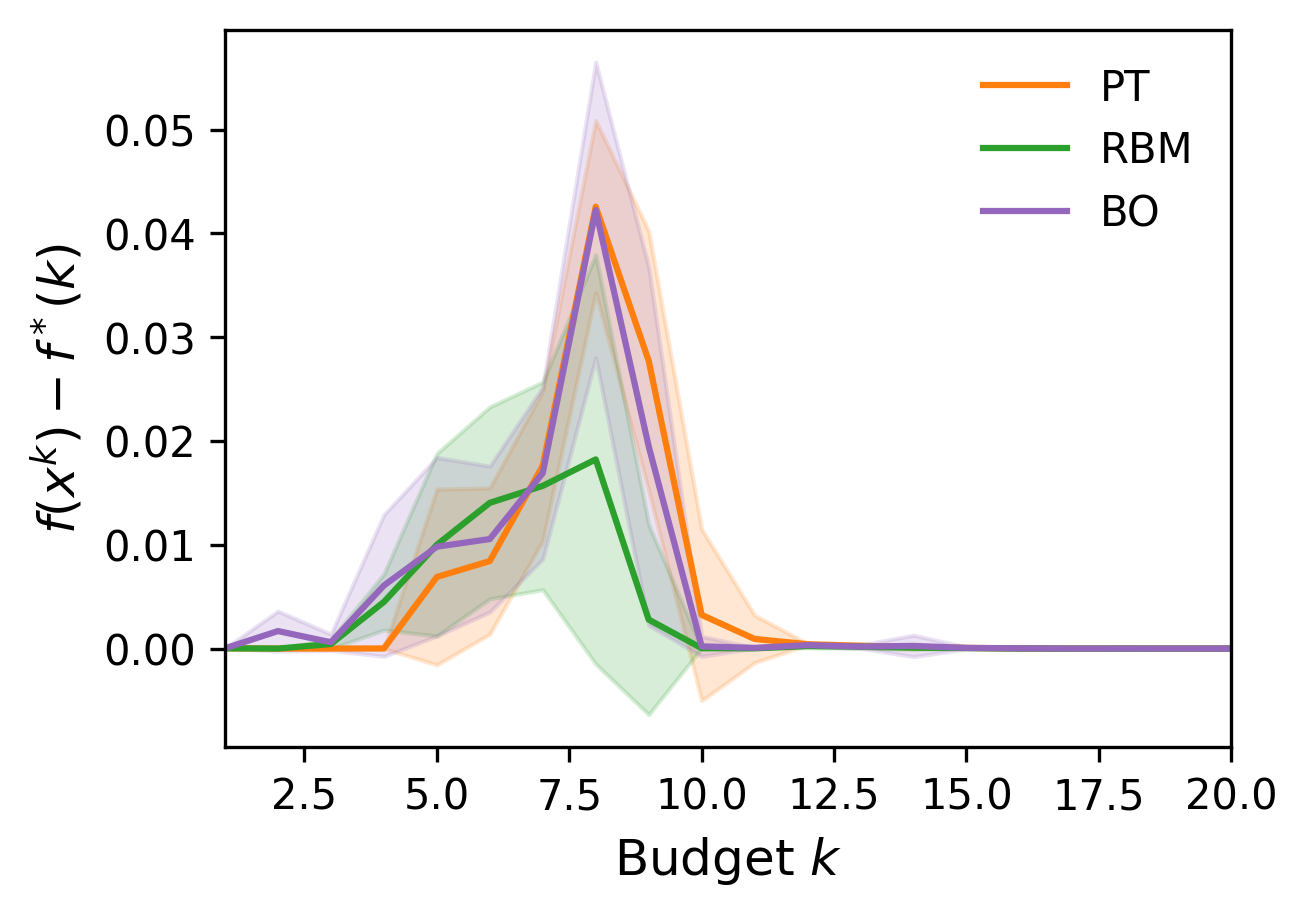}
        \phantomsubcaption\label{fig:log_distance_linear}
    \end{subfigure}
    \caption{\textbf{Distance to the ground-truth optimal path.}
    For each method and budget $k \in \{1,\ldots,20\}$, we plot
    $f(x^k_r) - f^*(k)$ across $N_\mathrm{rep}=30$ independent runs,
    where $f^*(k) \equiv \min_{x^k \in S_N^k} f(x^k)$ is the
    exhaustive-enumeration optimum at budget~$k$.
    Solid lines show the mean across runs; shaded bands show $\pm 1$
    standard deviation.
    \textbf{(a)}~Gap in $\log_{10}$ scale. Mean and standard deviation are
    computed on the log-transformed gaps, giving a spread that is symmetric
    and meaningful in log space.
    In the hard regime ($k \approx 7$--$13$), where $\binom{21}{k}$ peaks
    at $352\,716$, the RBM\,(green) reaches a mean log-gap of $-9.6$ at
    $k=11$ --- five orders of magnitude below PT\,(orange, $-4.2$) ---
    at one-tenth the SDP cost; RBM and BO\,(purple) converge to the
    numerical precision floor by $k\approx 17$, PT by $k=20$.
    \textbf{(b)}~Same gap on a linear scale. The dominant feature is a sharp
    peak near $k \approx 8$ where all methods are farthest from optimal,
    followed by rapid convergence to the precision floor by $k \approx 10$.
    The RBM peak is roughly half that of PT and BO; beyond $k = 10$ all three
    methods are indistinguishable from optimal on this scale.}
    \label{fig:log_distance}
\end{figure}

Figures~\ref{fig:comparison_runtime} and~\ref{fig:log_distance} compare the three methods in terms of
both solution quality and computational cost.  The left panel
(Fig.~\ref{fig:comparison}) shows that all methods track the optimal
curve closely throughout the transition region, with the RBM achieving a
substantially closer approach to optimal in the hard mid-range regime
(quantified in Fig.~\ref{fig:log_distance}), while BO converges at higher~$k$ but at a
substantially lower per-evaluation cost.  All three methods clearly outperform
the greedy baseline, confirming that individual-moment assessments are not adequate to navigate the non-modular landscape.

The linear-scale figures above compress differences once the methods approach
the optimal path, making precise quantitative comparisons difficult.
Figure~\ref{fig:log_distance} shows the gap $f(x^k) - f^*(k)$ in both
logarithmic and linear scales, giving complementary views of the same data.

Figure~\ref{fig:runtime} shows the normalized runtime of each method as a
function of $k$, expressed relative to the greedy algorithm and plotted on a
logarithmic scale.  The brute-force curve peaks at $\binom{21}{10} = 352\,716$
normalized evaluations at $k=10$ — several orders of magnitude above all three
methods across the full range.
BO maintains the lowest runtime throughout, remaining within one to two orders
of magnitude of the greedy baseline.  This reflects its fundamental design principle;
with $250$ SDP evaluations, the surrogate amortizes the cost of expensive
function calls, and the dominant overhead is model fitting and candidate
sampling rather than SDP evaluation.  The RBM occupies an intermediate
position, with a runtime that grows gradually with $k$ as the internal MCMC
steps become more expensive in the denser, higher-$k$ landscape.  PT incurs
the highest overhead due to its five-chain structure, which multiplies the
per-step evaluation cost by the number of replicas.
The steep rise in PT and RBM runtimes at large $k$ reflects a feature of the
problem geometry rather than a limitation of the methods: as $k$ approaches
$N$, the moment matrix becomes denser and each SDP solve is more expensive,
while the optimization itself becomes trivially easy.  Across the entire
problem range, all three methods remain orders of magnitude below the
brute-force cost, confirming the practical viability of the proposed framework.

The \textit{collective} behavior of moments determines how informative they are in tightening SDP bounds.
The synergy diagnostic $\Delta(S_k)$ tracks this
distinction reliably and, because it is computed from data already available
at step $k{-}1$, adds no extra SDP cost.
The choice among methods depends on the application. For problems where certification must be tight --- as in device-independent quantum key distribution~\cite{AcinEtAl2007} or claims of quantum advantage --- RBM is the better option: it reaches near-optimal bounds at costs that are orders of magnitude below exhaustive enumeration. When the goal is instead a fast survey of many scenarios, BO offers a useful trade-off: lower accuracy in the transition regime, but a factor of twenty fewer SDP evaluations than PT overall. PT is the natural choice when a physically motivated ansatz for a good moment subset is available: one chain can be initialized from the ansatz while the others explore the landscape from random configurations. This is precisely the strategy used in the warm-start optimization discussed in Sec.~\ref{sec:manybody}.

\subsection{Bell inequality use-case: testing (4,4,2,2) inequalities}
\label{sec:bell4422}
The $I_{3322}$ benchmark served its purpose precisely because
exhaustive enumeration makes the ground truth available at every $k$,
allowing a rigorous assessment of each method against the exact optimum.
The deeper question, however, is whether the rigid level structure of the NPA
hierarchy — in which one moves from \textsf{NPA1} to \textsf{NPA2} to
\textsf{NPA3} as discrete, analytically defined steps — is actually the most
natural unit of computational effort, or whether the physically relevant
quantity is instead the number of moments $k$ required to reach a given
accuracy within a fixed level window.
Framing the problem as a transition from $k=0$ to $k=N$ within the
$\textsf{NPA1} \to \textsf{NPA2}$ window suggests that different Bell
inequalities, even within the same scenario, may converge at very different
points along this transition. For some inequalities, a small fraction of the NPA2 moments may already suffice. For others however, most of the adding set may be necessary.
If this heterogeneity is systematic, it implies that the coarse
``\textsf{NPA2} bound'' conflates qualitatively different regimes of the
relaxation landscape, and that a moment-selective approach carries
operational meaning beyond the $I_{3322}$ case.

To test the generality of the transitional behavior between NPA levels, we apply the optimization framework developed in this work to the Bell inequalities in
the $(4,4,2,2)$ scenario, collected and analysed in Ref.~\cite{Cruzeiro2019}, whose \textsf{NPA2} bound serves as our convergence reference.
The adding set $\mathcal{A} = \textsf{NPA2} \setminus \textsf{NPA1}$
contains $N = 40$ moments, placing exhaustive enumeration entirely out of
reach ($\binom{40}{20} \approx 1.4 \times 10^{11}$ evaluations at the hardest
$k$).
Of the 174 inequalities in the scenario, we first discard those for which
$|f_\gamma^{\textsf{NPA1}} - f_\gamma^{\textsf{NPA2}}| < 10^{-4}$,
i.e.\ inequalities already effectively converged at \textsf{NPA1} and for
which the adding set carries no discriminating information.
This leaves $171$ non-trivial inequalities that form the basis of our
analysis.
For each of the 171 inequalities we run the RBM optimizer, which performed
best on the $I_{3322}$ benchmark — across the full range $k \in \{0, \ldots,
40\}$.  We then characterize convergence using two relative-error thresholds,
$\varepsilon = 5\%$ and $\varepsilon = 1\%$,
declaring convergence at the smallest $k^*$ satisfying
\begin{equation}
    \frac{|f_\gamma(S_{k^*}) - f_\gamma^{\textsf{NPA2}}|}
         {|f_\gamma^{\textsf{NPA2}} - f_\gamma^{\textsf{NPA1}}|}
    \leq \varepsilon.
\end{equation}

\begin{figure}[htbp]
    \centering
    \begin{subfigure}[b]{0.48\textwidth}
        \centering
        \includegraphics[width=\textwidth]{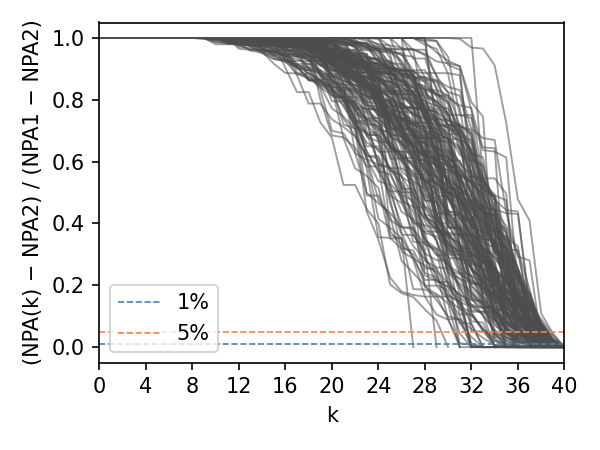}
        \phantomsubcaption\label{fig:bell4422_a}
    \end{subfigure}
    \hfill
    \begin{subfigure}[b]{0.48\textwidth}
        \centering
        \includegraphics[width=\textwidth]{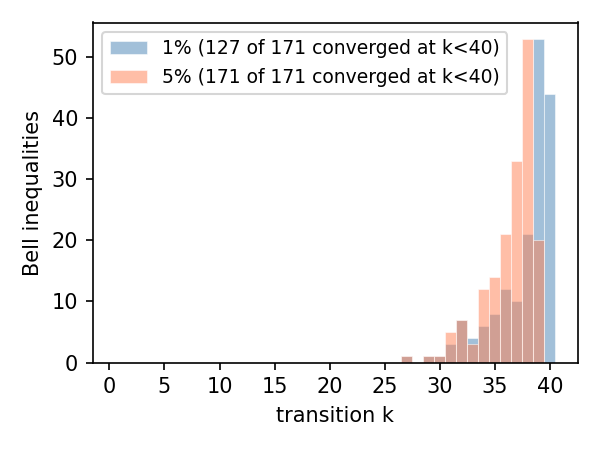}
        \phantomsubcaption\label{fig:bell4422_b}
    \end{subfigure}
    \caption{
    \textbf{Convergence analysis of 171 Bell inequalities in the
    $(4,4,2,2)$ scenario within the \textsf{NPA1}--\textsf{NPA2} window
    ($N=40$ moments).}
    Inequalities with
    $|f_\gamma^{\textsf{NPA1}} - f_\gamma^{\textsf{NPA2}}| < 10^{-4}$
    (already converged at \textsf{NPA1}) are excluded from the analysis.
    \textbf{(a)}~Normalised relaxation value as a function of Hamming weight
    $k$ for each of the 171 inequalities, obtained by running the RBM
    optimizer across the full range $k \in \{0,\ldots,40\}$.
    \textbf{(b)}~Distribution of convergence points $k^*$ under two
    relative-error thresholds: $5\%$ (orange) and $1\%$ (blue).
    All inequalities reach within $5\%$ of the \textsf{NPA2} value before the full adding set, while $1\%$ is reached by 127 of the 171 inequalities with $k$ strictly less than $40$.}
    \label{fig:bell4422}
\end{figure}

In Figure~\ref{fig:bell4422} we show the normalised relaxation value as a function of Hamming weight (a), as well as the convergence histograms at $\epsilon=1\%$ and $5\%$ (b). We find that all inequalities begin the transition from the \textsf{NPA1} value to the \textsf{NPA2} value between $k=9$ and $k=33$, most of them starting at $k\approx10-24$, well in the middle of the $k$ range. Around $k\approx 28-35$ most inequalities reach half of the gap in relative error. Finally, all inequalities reach within 5\% of their respective \textsf{NPA2} values \textit{before} k=40. To reach a 1\% relative error, for 127 inequalities it sufficed to use $k<40$, while for the remaining 44 running the full \textsf{NPA2} relaxation was necessary. 

A clear transitional behavior occurs between the levels \textsf{NPA1} and \textsf{NPA2} for all inequalities. From a practical point of view, our results confirm that if one does not have access to resources to run \textsf{NPA} level $M$, running heuristics such as the ones developed here are clearly beneficial rather than stopping at level $M-1$. The width of the transitional behaviors portrays key differences among different Bell-inequalities and  highlights the relevance of tailored moment-selection. The specific shape of the transition is related to the choice of the adding set, $\textsf{NPA2} \setminus \textsf{NPA1}$. An interesting future direction could be to examine how a larger adding set, like $\textsf{NPA3} \setminus \textsf{NPA1}$, could bring these transitions to even lower $k$ values, at a higher computational cost during exploration due to the larger state space. Our results indicate that in general it may be worth reconsidering the standard hierarchy, and to prefer a moment-selective approach. We provide a generic, algorithmic tool for this. 

\subsection{Application to many-body ground-state certification}
\label{sec:manybody}
The combinatorial moment-selection framework introduced in Sec.~\ref{sec:methods}
is formulated entirely in terms of the abstract objects $\mathcal{I}$,
$\mathcal{F}$, $\mathcal{A}$, and the SDP objective. It makes no assumption about
the physical interpretation of the moments, the structure of the polynomial
optimization, or the observable of interest.  This generality makes the framework applicable to any noncommutative polynomial optimisation problem, being the maximisation of Bell inequality violations one of the many possible instances.  

One natural application of the NPA relaxation approach consists of the certification of ground-state properties of quantum many-body systems~\cite{wang2024certifying}, where the same hierarchy provides certified bounds on observable expectation values.
We apply the framework to the one-dimensional spin-$\tfrac{1}{2}$ Heisenberg chain
with periodic boundary conditions,
\begin{equation}
\label{eq:heisenberg}
H = \sum_{i=1}^{N} \mathbf{S}_i \cdot \mathbf{S}_{i+1},
\end{equation}
with $\mathbf{S}_{N+1} \equiv \mathbf{S}_1$.  The exact ground-state energy per
site $e_0(N) = E_0(N)/N$ is known from the Bethe ansatz~\cite{bethe1931,hulthen1938},
providing an absolute benchmark against which the quality of any relaxation can
be measured precisely.  A key contribution of Ref.~\cite{wang2024certifying} is the
local basis $\mathcal{B}_\text{local}$, a physically motivated sparse monomial
set built from geometrically local Pauli strings --- single-site operators
$\sigma_i^a$, two-site correlators $\sigma_i^a\sigma_{i+j}^b$, and contiguous three-
and four-body strings --- with separations up to $r = \lceil N/2 \rceil$.
Locality naturally suits the energy, which depends only on nearest-neighbour
two-site correlators, and the local basis tracks the exact energy to within
$10^{-4}$--$10^{-2}$ across a wide range of system sizes.
From our perspective, $\mathcal{B}_\text{local}$ represents a hand-crafted instance
of the moment-selection problem: a fixed choice of $\mathcal{F}$ designed by
physical reasoning rather than by combinatorial optimization.  This prompts two questions.  First, is $\mathcal{B}_\text{local}$ \emph{internally compressible}, that is,
can a fixed fraction of its monomials reproduce most of its certification power
uniformly in system size?  Second, is it globally optimal at its size, or can
a broader candidate pool produce tighter certificates at the same computational budget?

\subsubsection{Fractional compression within the local basis}
\label{subsec:manybody_compression}
\begin{figure}[htbp]
    \centering
    \includegraphics[width=\textwidth]{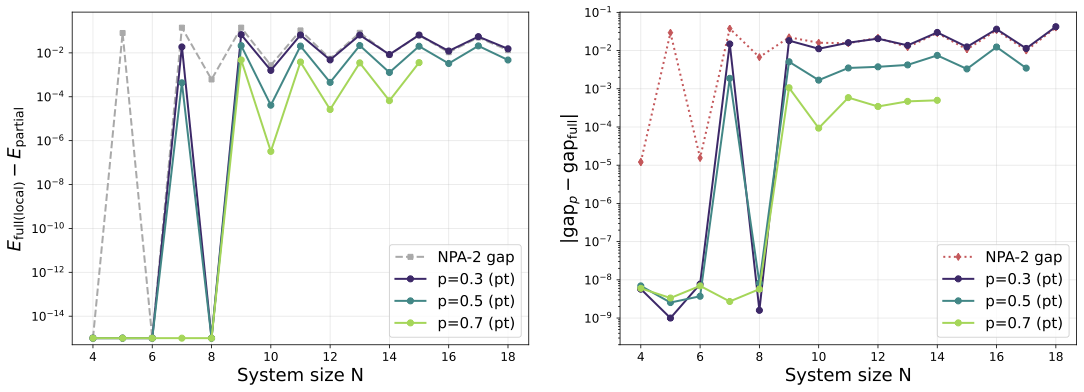}
    \caption{\textbf{Fractional compression of the local-basis relaxation for
    the 1D Heisenberg chain.}  For each system size $N$ and fraction
    $p \in \{0.3, 0.5, 0.7\}$, PT selects a subset of size
    $k = \lfloor p\,|\mathcal{A}(N)|\rfloor$ from the adding set
    $\mathcal{A} = \mathcal{B}_\text{local} \setminus \textsf{NPA1}$ via a
    warm-start sweep.  The dashed line shows the gap of the full \textsf{NPA2}
    truncation for comparison.
    \textbf{(a)}~Energy discrepancy $|E_\text{full}(N) - E_\text{partial}(N,p)|$
    relative to the full local-basis relaxation.  For $N \leq 8$ the fractions
    $p \geq 0.5$ reproduce the full relaxation to machine precision; for
    $N \gtrsim 9$ the curves stabilise to approximately $N$-independent plateaus
    on the even and odd subsequences separately, with $p = 0.7$ achieving errors
    in the $10^{-4}$--$10^{-5}$ range and $p = 0.3$ consistently below the full
    \textsf{NPA2} gap.  The even--odd splitting reflects frustration induced by
    periodic boundary conditions for odd $N$.
    \textbf{(b)}~Excess observable gap $\delta_C(N,p) = \Delta_C^{(p)}(N) -
    \Delta_C^\text{full}(N)$ on the half-chain correlator $C_{N/2}$.  Unlike the
    energy, the curves show persistent oscillations that do not stabilise with
    $N$: even $p = 0.7$ leaves a residual excess of order $10^{-3}$--$10^{-4}$
    at large system sizes, demonstrating that local-basis compression is
    intrinsically harder for long-range observables.}
    \label{fig:manybody_fractional}
\end{figure}

To probe internal compressibility, we set $\mathcal{I} = \textsf{NPA1}$ and
$\mathcal{F} = \mathcal{B}_\text{local}$, so that the adding set
$\mathcal{A} = \mathcal{B}_\text{local} \setminus \textsf{NPA1}$ contains all
local-basis monomials not already present at NPA level 1.  For each system size
$N$ and each target fraction $p \in \{0.3, 0.5, 0.7\}$, we set
$k(N,p) = \lfloor p\,|\mathcal{A}(N)|\rfloor$ and run PT over subsets
$\mathcal{S} \subseteq \mathcal{A}(N)$ of size $k$, using a warm-start sweep that
carries the best solution found at Hamming weight $k-1$ forward as the cold-replica
initialisation at weight $k$.
Figure~\ref{fig:manybody_fractional} presents the resulting discrepancy to the
full local-basis relaxation as a function of $N$, for both the ground-state energy
(panel~(a)) and the half-chain spin--spin correlator
$C_{N/2} = N^{-1}\sum_{i=1}^N \langle \mathbf{S}_i \cdot \mathbf{S}_{i+N/2}
\rangle$ (panel~(b)), alongside the gap of the full \textsf{NPA2} truncation
for comparison.

For the energy (panel~(a)), the first observation is that for small system sizes
$N \leq 8$ the optimised subsets at fractions $p = 0.5$ and $p = 0.7$ reproduce
the full local-basis relaxation to machine precision.  At these sizes the adding
set $\mathcal{A}(N)$ is small enough that a majority fraction of it already spans
the full pool, and PT trivially identifies the optimal choice.  The physically
interesting regime begins at $N \gtrsim 9$, where the adding set grows large
enough that the fraction $p$ becomes a genuine constraint.  From this point
onward, the error curves for each fixed $p$ stabilise to an approximately
$N$-independent plateau, separately on the even and odd subsequences.  The
parity splitting is a physical consequence of frustration: for odd $N$, the
antiferromagnetic ordering induced by the PBC cannot close consistently around
the ring, producing a geometrically frustrated ground state whose global character
is not captured by local SDP constraints.  Above this systematic splitting, the
central finding is clear: a fixed fraction of the local-basis monomials suffices
to approximate the full local-basis relaxation uniformly in system size.
Quantitatively, $p = 0.7$ achieves energy errors in the range $10^{-4}$--$10^{-5}$,
and even $p = 0.3$ consistently falls below the full \textsf{NPA2} gap for
$N \gtrsim 9$ despite using a much smaller monomial budget.  This directly extends
to the many-body setting the core result of Sec.~\ref{subsec:synergy_results}:
budget-aware selection outperforms naive level-based truncation.

The picture changes qualitatively for the half-chain correlator (panel~(b)).  Here
the discrepancy curves do not stabilise; they display oscillations whose amplitude
does not decrease with $N$, and even $p = 0.7$ shows excess gaps of order
$10^{-3}$--$10^{-4}$ above the full local-basis relaxation at large system sizes
--- values that remain significantly above machine precision and do not plateau.
$C_{N/2}$ probes spin--spin correlations at the maximum separation on the ring, yet the local basis is built exclusively
from Pauli strings supported on short contiguous intervals.  When the local pool is
subsampled, the operator products most relevant to constraining this nonlocal
observable are precisely those removed first.  Compressing the local-basis
relaxation is therefore intrinsically harder for nonlocal observables, and the
stable $N$-independent plateau that characterises the energy case does not emerge,
motivating the use of a fundamentally broader candidate pool.

\subsubsection{Optimality beyond the local basis}
\label{subsec:manybody_optimality}
\begin{figure}[htbp]
    \centering
    \begin{subfigure}[b]{0.48\textwidth}
        \centering
        \includegraphics[width=\textwidth]{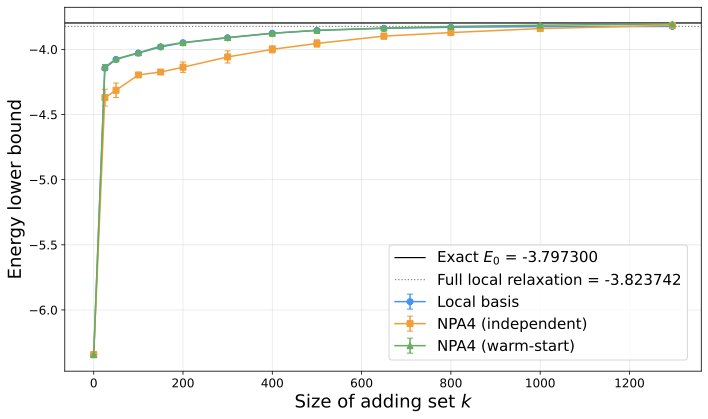}
        \phantomsubcaption\label{fig:energy_opt_full}
    \end{subfigure}
    \hfill
    \begin{subfigure}[b]{0.48\textwidth}
        \centering
        \includegraphics[width=\textwidth]{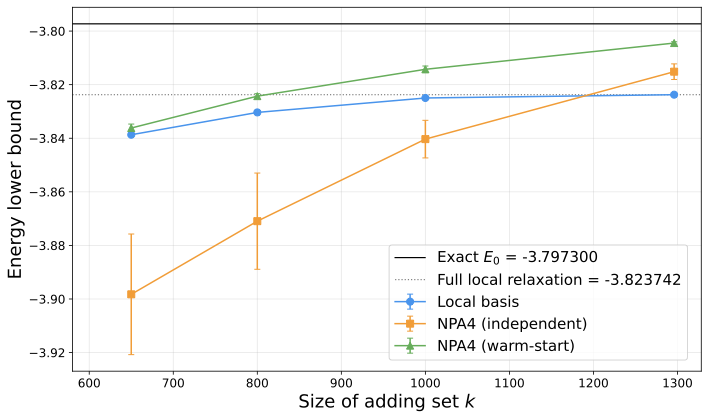}
        \phantomsubcaption\label{fig:energy_opt_zoom}
    \end{subfigure}
    \caption{\textbf{Energy optimality test beyond the local basis
    ($N = 9$, 1D Heisenberg chain).}  At matched monomial budget $k$, three PT
    strategies are compared: optimisation restricted to the local pool
    $\mathcal{A}_\text{local} = \mathcal{B}_\text{local} \setminus \textsf{NPA1}$
    (blue); optimisation over the enlarged pool
    $\mathcal{A}_\text{NPA4} = \textsf{NPA4} \setminus \textsf{NPA1}$ with
    standard warm-start (orange); and the same enlarged-pool search additionally
    seeded at each $k$ from the best local-pool solution (green, warm-started
    NPA4).  The solid line marks the exact ground-state energy
    $E_0 = -3.797300$; the dotted line marks the full local-basis bound
    $E_\text{full} = -3.823742$.  Error bars indicate run-to-run variability
    over independent PT initialisations.
    \textbf{(a)}~Full budget range.  The local-pool and warm-started NPA4
    curves rise steeply and converge near the local-basis bound, while the
    independent NPA4 search lags behind with larger variance.
    \textbf{(b)}~Zoom on the regime $k \geq 600$.  The warm-started NPA4 run
    (green) crosses the full local-basis bound around $k \approx 800$ and
    reaches $E_\text{LB} \approx -3.805$ at $k = 1285$, demonstrating that
    nonlocal NPA4 monomials can tighten the energy certificate beyond what any
    local-basis subset can achieve.  The independent NPA4 search (orange)
    exhibits large variance and remains below the local-basis bound across most
    of the budget range.}
    \label{fig:manybody_energy_opt}
\end{figure}

\begin{figure}[htbp]
    \centering
    \begin{subfigure}[b]{0.48\textwidth}
        \centering
        \includegraphics[width=\textwidth]{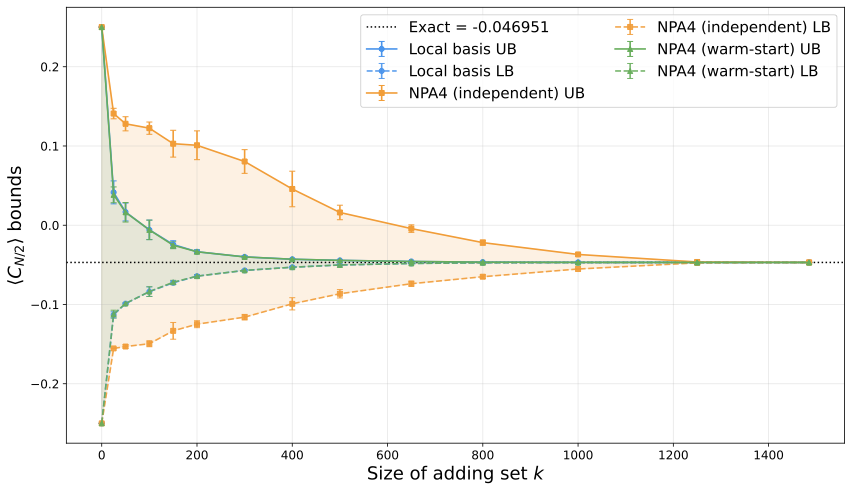}
        \phantomsubcaption\label{fig:manybody_obs_bounds}
    \end{subfigure}
    \hfill
    \begin{subfigure}[b]{0.48\textwidth}
        \centering
        \includegraphics[width=\textwidth]{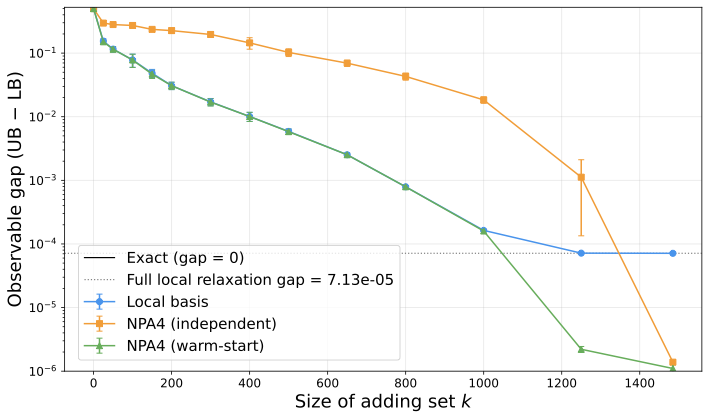}
        \phantomsubcaption\label{fig:manybody_obs_gap}
    \end{subfigure}
    \caption{\textbf{Observable optimality test beyond the local basis
    ($N = 10$, half-chain correlator $C_{N/2}$).}  The same three PT strategies
    as in Fig.~\ref{fig:manybody_energy_opt} are compared, now optimising the
    certified width of the interval on $\langle C_{N/2}\rangle_\text{gs}$.
    \emph{Left:} certified upper and lower bounds on $C_{N/2}$ as a function
    of $k$.  The exact value $C_{N/2}$ is shown as a dotted line.
    The shaded regions illustrate the width of the certified interval for each
    strategy.  The local-basis bounds (blue) are not visible in this panel
    because they are overlaid by the warm-started NPA4 bounds (green): the two
    strategies produce nearly indistinguishable intervals up to the point where
    the enlarged-pool search begins to improve beyond the local-basis ceiling,
    consistent with the warm-start initialisation.  The separation between the
    two becomes clearly visible in the right panel on the logarithmic scale.
    \emph{Right:} observable gap $\Delta_C = U_B - L_B$ on a logarithmic
    scale.  The local-pool search (blue) saturates at the full local-basis gap
    of $7.13 \times 10^{-5}$ (dotted line).  The warm-started NPA4 run (green)
    crosses this level around $k \approx 1000$ and reaches
    $\Delta_C \approx 10^{-6}$ at $k \approx 1300$, a reduction of nearly two
    orders of magnitude, demonstrating that tight certification of long-range
    correlations requires genuinely nonlocal monomials unavailable within
    $\mathcal{B}_\text{local}$.  The independent NPA4 search (orange) barely
    reaches the local-basis gap within the tested budget.}
    \label{fig:manybody_obs_opt}
\end{figure}

We next ask whether $\mathcal{B}_\text{local}$, the moment selection of~\cite{wang2024certifying}, is globally optimal at its size,
or whether monomials outside the local construction --- higher-degree or
geometrically non-contiguous Pauli strings --- can produce tighter certificates
at the same budget.  To answer this, we enlarge the candidate pool to
$\mathcal{F} = \textsf{NPA4}$ while keeping $\mathcal{I} = \textsf{NPA1}$, and
compare, at matched Hamming weight $k$, three optimisation strategies: (i) PT
restricted to the local pool
$\mathcal{A}_\text{local} = \mathcal{B}_\text{local} \setminus \textsf{NPA1}$;
(ii) PT over the full enlarged pool
$\mathcal{A}_\text{NPA4} = \textsf{NPA4} \setminus \textsf{NPA1}$ with a standard
$k$-to-$k$ warm-start; and (iii) the same enlarged-pool PT additionally seeded, at
each $k$, with the best solution found by the local-pool search at the same budget
--- hereafter the \emph{warm-started} NPA4 run.  The third strategy embeds the
physical prior encoded by $\mathcal{B}_\text{local}$ as a structured initialisation
for the enlarged search, rather than treating the two pools independently.
Figure~\ref{fig:manybody_energy_opt} reports the energy lower bound as a function
of $k$ for $N = 9$: panel~(a) shows the full budget range, while panel~(b) zooms
on the large-$k$ regime $k \geq 600$ where the three strategies separate.
Figure~\ref{fig:manybody_obs_opt} shows the corresponding results for the
half-chain correlator $C_{N/2}$ at $N = 10$: the left panel displays the certified
upper and lower bounds on $\langle C_{N/2}\rangle_\text{gs}$, and the right panel
reports the observable gap $\Delta_C = U_B(C_{N/2}) - L_B(C_{N/2})$ on a
logarithmic scale.  For the observable, the SDP is formulated with an
energy-window constraint that restricts the feasible set to states consistent with
the energy relaxation, so that the tightness of the observable certificate directly
reflects the monomial structure of the relaxation.

The energy results (Fig.~\ref{fig:manybody_energy_opt}) reveal a clear ordering
across the full budget range (panel~(a)): the independent NPA4 search lags behind
the local-pool optimisation, with large run-to-run variance that persists up to
$k \approx 1000$.  The local pool (blue) saturates at the full local-basis bound
(dotted line, $E_\text{full} = -3.823742$) and does not improve further.  The
zoomed view (panel~(b)) resolves the large-$k$ regime: the warm-started NPA4 run
(green) crosses the full local-basis bound around $k \approx 800$ and continues
to improve, reaching $E_\text{LB} \approx -3.805$ at the maximum tested budget
$k = 1285$, compared to the exact energy $E_0 = -3.797300$ (solid line).
This demonstrates that nonlocal NPA4 monomials can tighten the energy certificate
beyond what any subset of $\mathcal{B}_\text{local}$ can achieve, provided the
enlarged-pool search is initialised from the local-basis solution.

Figure~\ref{fig:manybody_obs_opt} shows a qualitatively stronger effect for the
half-chain correlator.  The local-pool optimisation (blue) tightens $\Delta_C$
rapidly and saturates at the full local-basis gap of $7.13 \times 10^{-5}$
(dotted line in the right panel).  The independent NPA4 search (orange) remains
far above this level throughout most of the budget range --- with
$\Delta_C \gtrsim 10^{-2}$ up to $k \approx 1200$ --- and only begins to
approach the local-basis level near the maximum tested budget $k \approx 1300$,
reflecting the difficulty of navigating the much larger landscape
$\mathcal{A}_\text{NPA4}$ without a structured initialisation.  The warm-started
NPA4 run (green), by contrast, crosses the full local-basis gap around
$k \approx 1000$ and continues to decrease, reaching $\Delta_C \approx 10^{-6}$
at $k \approx 1300$ --- an improvement of nearly two orders of magnitude over the
full local-basis relaxation.  This dramatic gain reflects the structural
mismatch between the local basis and the observable: certifying $C_{N/2}$
requires constraints arising from genuinely long-range operator products that
$\mathcal{B}_\text{local}$ cannot provide by construction, and once these monomials
are accessible in the enlarged pool, the warm-started optimizer identifies and
exploits them effectively.

The moment-selection framework
introduced in this work applies naturally and productively to the many-body
certification setting.  The local basis of Ref.~\cite{wang2024certifying} is a
highly effective hand-crafted heuristic for local observables, but it is not
globally optimal, and for observables probing long-range correlations the certified
bound can be improved by nearly two orders of magnitude using a budget-aware search
over a broader candidate pool.  The warm-start strategy --- using the local-basis
solution as a structured initialisation for the enlarged-pool search --- is
essential for navigating the much larger combinatorial landscape efficiently, and
provides a principled and practical recipe for extending high-quality ground-state
certification to system sizes and observables where rigid NPA level truncations are
provably suboptimal.

\section{Conclusion}
\label{sec:conclusion}

We have studied the problem of selecting which moments to include in
NPA relaxations for noncommutative polynomial optimisation when computational resources are limited.  The
results span different domains of increasing complexity and together establish both
the structural richness of the moment-selection landscape and the practical
effectiveness of the optimization framework developed here.

Working on the $I_{3322}$ benchmark, where exhaustive enumeration over all
$2^{21}$ subsets provides ground truth at every budget~$k$, we find that the
improvement in the relaxation bound is not distributed uniformly across subset
sizes, but is concentrated in a sharp transition window $5 < k \leq 19$.
Within this window, subsets of identical size can yield bounds spanning nearly
the full range from the \textsf{NPA1} value to the \textsf{NPA2} optimum.  The
landscape is highly heterogeneous in this regime: the typical value across random
subsets is far from the best achievable value.  The root cause is strong
higher-order synergy among moments --- the optimal subset of size~$k$ cannot be
built by extending the optimal subset of size $k{-}1$ --- which causes greedy
strategies to fail entirely in the transition regime.

The marginal synergy $\Delta(S_k)$ provides a useful convergence diagnostic at
no additional computational cost.  In the transition regime, the standard
marginal cost improvement is nearly zero even though large gains remain
accessible at higher~$k$; the synergy, which measures how much the bound
degrades when any single moment is removed from the current best subset, stays
elevated throughout and decays only once the landscape becomes approximately
modular.  In practice, monitoring $\Delta(S_k)$ alongside the cost curve
guards reliably against premature stopping.

Beyond brute-force approaches, three optimization strategies were introduced --- Parallel Tempering, a
gradient-trained Restricted Boltzmann Machine, and Bayesian Optimization with a
random-forest surrogate ---  and tested. They all outperform the greedy baseline by a wide margin, at
costs around two orders of magnitude below brute force.  PT is
competitive at low budget; the RBM dominates in the hard transition regime,
approaching the optimum up to five orders of magnitude more closely than PT
despite a tenfold smaller evaluation budget; and both converge at high budget.
BO trades accuracy in the transition regime for a ${\sim}20$-fold reduction in
SDP evaluations relative to PT, making it preferable when evaluations are the
dominant cost.  The method that performs best --- RBM --- is
precisely the one whose design facilitates collective exploration, consistent with
the synergistic character of the landscape.

Applying the framework to all 174 Bell inequalities in the $(4,4,2,2)$ scenario,
where the adding set grows to $N = 40$ moments and exhaustive search is
infeasible, we find substantial heterogeneity in the moment budget $k^*$
required to reach a given fraction of the \textsf{NPA1}--\textsf{NPA2} window:
some inequalities converge with only a small fraction of the \textsf{NPA2}
adding set, while others require nearly the full pool.  This confirms that the
coarse label ``\textsf{NPA2} bound'' conflates qualitatively different
convergence regimes, and that a moment-selective approach carries genuine
operational meaning across an entire Bell scenario.

Extending the framework to the one-dimensional Heisenberg spin chain, where NPA
relaxations provide certified bounds on energies and observable expectation
values, yields two results.  First, the physically motivated local basis of
Ref.~\cite{wang2024certifying} is internally compressible for the ground-state
energy: a fixed fraction $p$ of its monomials reproduces the full local-basis
relaxation to an approximately $N$-independent accuracy, with even $p = 0.3$
outperforming the full \textsf{NPA2} truncation at most system sizes.  For the
half-chain correlator $C_{N/2}$, by contrast, the compression property breaks
down, as the error does not plateau with $N$ --- reflecting the fundamental
mismatch between a locally constructed basis and a nonlocal observable.
Second, the local basis is not globally optimal at its size: enlarging the
candidate pool to \textsf{NPA4} and warm-starting from the local-basis solution
improves the certified observable gap on $C_{N/2}$ by nearly two orders of
magnitude, demonstrating that genuinely nonlocal monomials are essential for
tight certification of long-range correlations.

Looking ahead, the most pressing open question is theoretical: which properties of a noncommutative polynomial optimization — the degree of the polynomial, the locality of the algebraic relations, the geometry of the feasible set — determine whether a synergy-dominated regime exists and how sharp the transition is. Knowing this would let one anticipate landscape difficulty before any SDP is solved. On the algorithmic side, the natural next step is to turn $\Delta(S_k)$ from a passive diagnostic into an active guide: rather than adding all moments at the next NPA level uniformly, one could grow the hierarchy selectively in the directions the synergy identifies as most structurally important, making NPA relaxations themselves budget-aware. We expect synergy-dominated landscapes to be the rule rather than the exception whenever the objective encodes the collective behaviour of many interacting degrees of freedom — a situation that arises across steady-state certification~\cite{mortimer2025certifying}, dimension witnesses, and many-body problems beyond the Heisenberg chain — precisely the settings where higher NPA levels are most needed and expensive.

\section*{Code availability}
The code implementing the moment optimization and reproducing all
numerical results is available at \url{https://github.com/Ciscone7/OptimalMomentsNPA}.

\section*{Acknowledgments} F.F is fellow of Eurecat’s “Vicente López”
Ph.D. grant program. The authors acknowledge financial support from the Government of Spain (Severo Ochoa CEX2019-000910-S and FUNQIP), Fundació Cellex, Fundació Mir-Puig, Generalitat de Catalunya (CERCA program). T.K. additionally acknowledges funding from the Swiss National Science Foundation (projects 214458 and 235415)

\bibliographystyle{plain}
\bibliography{references}
\end{document}